\documentclass[12pt]{iopart}

\usepackage{citesort}

\usepackage{graphicx}
\usepackage{bm}
\usepackage{subfigure}

\usepackage{color}

\begin{document}

\title[On the relationship between Gaussian stochastic blockmodels and label propagation algorithms]{On the relationship between Gaussian stochastic blockmodels and label propagation algorithms}

\author{Junhao Zhang$^1$\footnote{Co-corresponding author.}, Tongfei Chen$^2$\footnote{Performed while studied at Peking University.}, Junfeng Hu$^1$\footnote{Corresponding author.} }

\address{1 Key Laboratory of Computational Linguistics (Ministry of Education), Peking University, Beijing, P. R. China}

\address{2 Center for Language and Speech Processing, Johns Hopkins University, Baltimore, MD, USA}
\eads{\mailto{junhao.zhang@pku.edu.cn}, \mailto{tongfei@jhu.edu}, \mailto{hujf@pku.edu.cn}}

%\vspace{10pt}
%\begin{indented}
%\item[]%September 2014
%\end{indented}

\begin{abstract}
The problem of community detection receives great attention in recent years.
Many methods have been proposed to discover communities in networks.
In this paper, we propose a Gaussian stochastic blockmodel that uses Gaussian distributions to fit weight of edges in networks for non-overlapping community detection.
%The maximum likelihood estimate of this model is equivalent to label propagation with node preference.
The maximum likelihood estimation of this model has the same objective function as general label propagation with node preference.
The node preference of a specific vertex turns out to be a value proportional to the intra-community eigenvector centrality (the corresponding entry in principal eigenvector of the adjacency matrix of the subgraph inside that vertex's community) under maximum likelihood estimation. Additionally, the maximum likelihood estimation of a constrained version of our model is highly related to another extension of label propagation algorithm, namely, the label propagation algorithm under constraint.
Experiments show that the proposed Gaussian stochastic blockmodel performs well on various benchmark networks.
\end{abstract}

\vspace{2pc}
\noindent{\it Keywords}: random graphs, networks, clustering techniques, analysis of algorithms
%\maketitle

\section{Introduction}\label{introduction}

In complex networks, communities \cite{girvan2002community} are considered as groups of vertices whose intra-connections are dense and inter-connections are sparse. Many real-world networks contain community structures. Detection of communities in networks has been developed a lot in recent years.

Since the proposal of modularity \cite{newman2004finding} as a quality function of a community partition, many algorithms were proposed to maximize modularity. Meanwhile, label propagation algorithms \cite{raghavan2007near} are much simpler and more scalable for very large networks. Label propagation algorithm has many extensions, such as balanced label propagation algorithm \cite{vsubelj2011robust}, label propagation under constraint \cite{barber2009detecting} and label propagation with hop attenuation and node preference \cite{leung2009towards,vsubelj2011unfolding}. Barber and Clark \cite{barber2009detecting} showed that modularity can be viewed as a special case of label propagation under constraint.
Liu and Murata \cite{liu2010advanced} introduced multi-step greedy agglomeration \cite{schuetz2008efficient} to the label propagation approach used by Barber and Clark \cite{barber2009detecting}, which helps to escape from poor local optimum.
Leung et al. \cite{leung2009towards} first introduced hop attenuation and node preference to label propagation algorithm. \v{S}ubelj and Bajec \cite{vsubelj2011unfolding} used a formulation similar to PageRank \cite{brin1998anatomy} inside each community as the node preference and use the defensive diffusion and attenuation label propagation algorithm to extract the cores of communities.

Recent years also see the popularity of stochastic blockmodels \cite{holland1983stochastic,hofman2008bayesian,karrer2011stochastic,airoldi2008mixed,aicher2014learning,gopalan2013modeling} in both non-overlapping and overlapping community detection. Stochastic blockmodels assume that vertices in a network are partitioned into blocks, and edges of two vertices are sampled from a distribution parameterized by block indicators of two vertices. Stochastic blockmodels are flexible and can not only handle community detection, but also group vertices with similar behaviors, that is, vertices from one block have similar connectivity patterns with vertices from another block. For community detection problem, it implies dense intra-community connections and sparse inter-community connections.

Karrer and Newman \cite{karrer2011stochastic} proposed degree-corrected stochastic blockmodel which incorporates vertex degree into stochastic blockmodels and found that degree-corrected stochastic blockmodels perform better than standard stochastic blockmodels in networks with substantial heterogeneous degrees. Aicher et al. \cite{aicher2013adapting} and Aicher et al. \cite{aicher2014learning} proposed the weighted stochastic blockmodel to model weighted networks.

However, in community detection literature, there is little work on the nature of label propagation algorithm and label propagation with node preference. Tib{\'e}ly et al. \cite{tibely2008equivalence} proposed that the objective function of label propagation algorithm \cite{raghavan2007near} is equivalent to a ferromagnetic Potts model. Barber and Clark \cite{barber2009detecting} extended the objective function to a constrained one.
For label propagation with node preference \cite{leung2009towards,vsubelj2011unfolding}, the definition of node preference is still heuristic. In this paper, we establish a connection between stochastic blockmodels and label propagation and some of its extensions, so that new node preference parameters are derived in a mathematically rigorous framework, namely, through maximum likelihood estimation.

We propose the Gaussian stochastic blockmodel with node preference (GSBM-P) for non-overlapping community detection. Our model can be viewed as a constrained version of degree corrected version of the weighted stochastic blockmodel \cite{aicher2013adapting}. GSBM-P uses Gaussian distributions to model the weight of edges with the same global variance shared across all communities. It is assumed that the expected weight of edges between two different communities should be zero. GSBM-P assumes that the weight of an edge inside a community is determined by the node preference of its two vertices.
%The maximum likelihood estimation of GSBM-P turns out to be equivalent to label propagation with node preference.
The maximum likelihood estimation of GSBM-P turns out to have the same objective function as general label propagation with node preference.
The vector whose entries are node preferences of vertices inside a community is proportional to the $L_2$-normalized principal eigenvector of the adjacency submatrix computed inside the community. Additionally, the norm of this vector turns out to be the square root of the principal eigenvalue of this adjacency submatrix. This node preference as a local metric (i.e., a value proportional to intra-community eigenvector centrality \cite{bonacich1987power}) has never been proposed before. We show that the objective function of our model is equivalent to the sum of the squares of the principal eigenvalues of the adjacency matrices of all communities. Also, we show that the objective function of label propagation under constraint \cite{barber2009detecting} is a special case of the objective function of the maximum likelihood estimation of a constrained version of our model.

A coordinate ascent algorithm is developed to optimize the objective function of our model. We experimented on two real-world networks and also synthetic networks, and illustrated that our method performs better than label propagation with leading eigenvector of intra-community random walk matrix as node preference \cite{vsubelj2011unfolding} and the variational Bayes method for weighted stochastic blockmodel \cite{aicher2014learning} in community detection. Our method is also compared with label propagation algorithm (LPA) \cite{raghavan2007near} and order statistics local optimization method (OSLOM) \cite{lancichinetti2011finding}. In synthetic networks, our method achieves a performance superior or comparable to that of LPA. In unweighted synthetic networks, the performance of our method is close to that of OSLOM.

The rest of the paper is organized as follows: Section \ref{SectionRelatedWork} reviews related work; Section \ref{SectionGWSBM} introduces our model and the maximum likelihood estimation of our model; Section \ref{SectionExperiments} presents our experiments and the last section concludes this paper.

\section{Related Work}\label{SectionRelatedWork}

\subsection{Label Propagation Algorithm}

Label propagation algorithm \cite{raghavan2007near} seeks to discover communities by propagating vertices' labels. It iteratively updates a vertex's label by choosing the label possessed most by its neighbors. If we denote the label of vertex $i$ as $z_i$, then the update step of $z_i$ can be formulated as
\begin{equation}
z_i = \arg\max_z \sum_{j \neq i} W_{ij} \delta_{z_j z},
\end{equation}
where $W$ is the adjacency matrix, and $\delta_{ab} = 1$ iff $a = b$ is the Kronecker delta function. When the iteration process terminates, all vertices with the same labels are considered as a community.

Tib{\'e}ly et al. \cite{tibely2008equivalence}, and also Barber and Clark \cite{barber2009detecting} showed that the objective function of label propagation algorithm can be formulated as
\begin{equation}\label{lpaQ}
Q_{\mathrm{LPA}} = \sum_{i,j} W_{ij} \delta_{z_i z_j} \ = \sum_z \sum_{i,j} W_{ij} \delta_{z_i z} \delta_{z_j z} \ .
\end{equation}
This objective function is the total weight of the intra-community edges.
It is shown that label propagation algorithm aims to maximize $Q_{\mathrm{LPA}}$ through a zero-temperature kinetics. It can be seen that this objective function encourages large communities. Thus, label propagation algorithm sometimes gets a trivial solution where all vertices are in one community and the objective function reaches its maximum.

\subsection{Label Propagation with Node Preference}\label{DPA}

Leung et al. \cite{leung2009towards} introduced hop attenuation and node preference to label propagation algorithm.
We leave out the introduction of hop attenuation regularization which prevents the formation of very large communities.
They proposed a label update strategy
\begin{equation}
z_i = \arg\max_z \sum_{j \neq i} f_j^m W_{ij} \delta_{z_j z}  \ ,
\end{equation}
where $f_j$ is the propagation strength (a.k.a. node preference) of node $j$, $m$ is a parameter which controls the influence of $f_j$.

\v{S}ubelj and Bajec \cite{vsubelj2011unfolding} proposed the defensive diffusion and attenuation label propagation algorithm to accurately detect community cores. They define the propagation strength $f_i^m$ of node $i$ in community $z_i$ as the probability of a random walker inside the community labeled with $z_i$ visits vertex $i$, i.e.,
\begin{equation}
p_i^{z_i} = \sum_{j \in N(i)} \frac{p_j^{z_j}}{k_j^{z_i}} \delta_{z_i z_j}  \ ,
\end{equation}
where $N(i)$ denotes the set of vertices which are the neighbors of vertex $i$, and $k_j^{z_i}$ denotes the total weight of edges which connect vertex $j$ and vertices from community $z_i$. Note that this definition of node preference is leading eigenvector of random walk matrix inside the community $z_i$ and can be easily extended to weighted networks. In the following part of this paper, we shall call this value as intra-community RandomWalk or local RandomWalk.

Then, defensive diffusion applies preference to the core of each community,
\begin{equation}
z_i = \arg\max_z \sum_{j \neq i} p_j^{z_j} W_{ij} \delta_{z_j z} \ .
\end{equation}

However, the label propagation with intra-community RandomWalk as node preference prefers small communities, and normally discovers much more communities than what is expected.

Generally, the objective function that label propagation with node preference tries to maximize can be formulated as
\begin{equation}\label{Q_LPA-P}
Q_\mathrm{LPA-P} = \sum_z \sum_{i,j} p_i p_j W_{ij} \delta_{z_i z} \delta_{z_j z}  \ ,
\end{equation}
where $p_i$ is the node preference associated with vertex $i$, and it can be node degree or other centrality not depending on vertex's label.

We propose the Gaussian stochastic blockmodel with node preference (GSBM-P), whose maximum likelihood estimation results to the same objective function as general label propagation with node preference, where the node preference of a vertex is proportional to local eigenvector centrality computed inside its community rather than the local RandomWalk score. In other words, label propagation with intra-community RandomWalk as node preference does not have this objective function since the objective function may decrease once node preference is re-estimated. An alternative measure for eigenvector centrality  \cite{bonacich1987power} in directed network is the authority score and hub score in HITS \cite{kleinberg1999authoritative}. In the field of ontology learning, He et al. \cite{he2014construction} used an HITS based community detection algorithm to produce a concept hierarchy.

In this paper, we focus on undirected networks, and the proportion of intra-community eigenvector centrality is derived from maximum likelihood estimation.

\subsection{Weighted Stochastic Blockmodel}

Aicher et al. \cite{aicher2013adapting} and Aicher et al. \cite{aicher2014learning} applied stochastic blockmodel to weighted networks. For dense networks, they claimed that Gaussian distributions can be used to generate the weights of the edges between each pair of vertices. That is,
\begin{equation}
W_{ij} \sim \mathcal{N}(\mu_{z_i z_j}, \sigma_{z_i z_j}^2) \;.
\end{equation}

For sparse networks, they claimed that absent edges are different from edges with zero weight. Therefore, they modelled the edge existence and edge weights separately. Exponential family distribution such as Gaussian distribution is adopted only for weighted edges. They further showed that degree-corrected stochastic blockmodel \cite{karrer2011stochastic} can be adopted to model the existence of edges in the following manner:
\begin{equation}\label{WSBM}
\begin{array}{l}
\log P(W|{\bf{z}},{\bf{p}},{\bm{\theta }},{\bm{\mu }},{\bm{\sigma }})\\
 = \alpha \displaystyle\sum_{ij} \log \mathcal{P}({A_{ij}}|{p_i}{p_j}{\theta _{{z_i},{z_j}}})+ (1 - \alpha )\displaystyle\sum_{(i,j) \in E} \log \mathcal{N}({W_{ij}}|{\mu _{{z_i}{z_j}}},\sigma _{{z_i}{z_j}}^2)\:,
\end{array}
\end{equation}
where $\mathcal{P}$ denotes the distribution for modeling edge existence, $A_{ij}$ is 1 when vertex $i$ and $j$ is connected and 0 otherwise, $(i,j) \in E$ denotes a pair of linked vertices, $p_i$ is the parameter associated with vertex $i$ and ${p_i}{p_j}{\theta _{{z_i},{z_j}}}$ is the expectation of $A_{ij}$, and $\alpha$ is between 0 and 1. Note that the authors did not propose a degree-corrected version of weighted stochastic blockmodel for modeling edge weights.

However, the maximum likelihood estimation of $\sigma_{i,j}^2$ will be zero if the weights of all the edges between group $i$ and group $j$ are the same, which creates a degeneracy in the likelihood calculation. Therefore, a Bayesian approach is adopted. Prior distributions are introduced to all parameters and variable $\mathbf{z}$ as well.

We propose a Gaussian stochastic blockmodel that can be viewed as a constrained version of degree-corrected version of weighted stochastic blockmodel. It assumes blocks are assortative in the way that it explicitly assumes sparse inter-block connections, so that it is more suitable for community detection.

\section{Model}\label{SectionGWSBM}

In our model, namely Gaussian stochastic blockmodel with node preference(GSBM-P), it is assumed that the graph is generated by
\begin{itemize}
\item Assigning block indicators to each vertex;
 \item Then drawing edge weights from a Gaussian distribution for each edge (including non-existing edges with weight equal to zero) between each pair of vertices.
\end{itemize}

The Gaussian distribution for the weight of the edge between each pair of vertices is defined as follows:
\begin{equation}
W_{ij} \sim \mathcal{N}(p_i p_j \delta_{z_i z_j}, \sigma^2) \;.
\end{equation}
where $p_i$ is the node preference of vertex $i$, $z_i$ is the block indicator for vertex $i$, and $\sigma^2$ is the variance of Gaussian distributions.

This model assumes that the expected value of the weights of inter-community edges is 0; the expected value of the weights of the intra-community edges is a product of the node preference of the two vertices associated with the edges.

Our model can be viewed as a constrained version of the degree corrected version of weighted stochastic blockmodel(WSBM) proposed by Aicher et al. \cite{aicher2013adapting}. We extend the WSBM \cite{aicher2013adapting} to be degree corrected when explaining edge weights, rather than incorporate the degree corrected SBM \cite{karrer2011stochastic} to explaining edge existence, which is proposed in \cite{aicher2014learning}. The degree correction should make our model fit better than WSBM in networks with substantial heterogeneous degrees (or sum of linked edge weights). Meanwhile, our model can be applied in assortative networks. By explicitly assuming the connection between blocks should be sparse, our model can discover community structures accurately.

\subsection{The Maximum Likelihood Estimation of GSBM-P}

In this section, we adopt the maximum likelihood estimation to fit our model, and show its objective function is the same as general label propagation with node preference. Since our model does not introduce priors over block indicators $\mathbf{z}$, we treat the block indicators as parameters, and use coordinate ascent method to estimate both the block indicators and node preference $\mathbf{p}$. The adopted coordinate ascent method estimates a parameter to maximize the likelihood when fixing all other parameters.

The likelihood function of our model is
\begin{equation}
P(W|\mathbf{z},\mathbf{p},\sigma)=\prod_{i,j}\frac{1}{\sigma\sqrt{2\pi}} \exp{\frac{ -\left( W_{ij}-p_i p_j \delta_{z_i z_j} \right)^2 }{2\sigma^2}} \ ,
\end{equation}
where $p_i$ is node preference of vertex $i$.

Maximizing the likelihood function is equivalent to minimizing $\sigma$ for Gaussian models.
Maximum likelihood estimation of $\sigma^2$ yields
\begin{equation}\label{sigma_model2}
\sigma^2=\frac{\sum_{i,j}\left(W_{ij}-p_i p_j \delta_{z_i,z_j}\right)^2}{n^2} \ .
\end{equation}
where $n$ is the total number of vertices.

Minimizing $\sigma^2$ is equivalent to maximizing the following objective function:
\begin{equation}\label{GSBM-P-OQ}
\begin{array}{rcl}
Q_\mathrm{GSBM-P} &=& \displaystyle 2\sum_{i,j} p_i p_j W_{ij} \delta_{z_i z_j} - \sum_{i,j} (p_i p_j \delta_{z_i z_j})^2 \\
&=& \displaystyle 2\sum_z \sum_{i,j} p_i p_j W_{ij} \delta_{z_i z} \delta_{z_j z} - \sum_z \sum_{i,j} (p_i p_j \delta_{z_i z} \delta_{z_j z})^2 \;.
\end{array}
\end{equation}

The maximum-likelihood estimated value of $p_i$ can be expressed in the following iterative updating manner:
\begin{equation}\label{p_update}
p_i^{(t+1)} =\frac{ \sum_j p_j^{(t)} W_{ij}\delta_{z_j z_i}}{\sum_j \left( p_j^{(t)}  \right)^2 \delta_{z_j z_i}} \;,
\end{equation}
or in vector and matrix notation:
\begin{equation}
\mathbf{p}_z^{(t+1)} = \frac{\mathbf{W}_z \mathbf{p}_z^{(t)}}{ \left\| \mathbf{p}_z^{(t)} \right\| ^2 } \;,
\end{equation}
where $\mathbf{p}_z$ is a vector whose entries consist of the node preferences of the vertices inside community $z$; $\mathbf{W}_z$ is the adjacency matrix of community $z$, that is, $\mathbf{W}_z$ describes how vertices inside community $z$ connected to each other while ignores the links outside community $z$.

Vector $\mathbf{p}_z$ converges to the principal eigenvector of $\mathbf{W}_{z}$. It can be shown that the norm of $\mathbf{p}_z$ is the square root of the principal eigenvalue of $\mathbf{W}_{z}$, denoted as $\sqrt{\lambda_z}$. Thus, these node preferences are the product of $\sqrt{\lambda_z}$ and the $L_2$-normalized local eigenvector centrality computed inside the community $z$.

Noticing that for a specific community $z$,
\begin{equation}\label{observation_1}
\sum_{i,j} p_i p_j W_{ij}\delta_{z_i z}\delta_{z_j z}=\mathbf{p}_z^\mathrm{T} \mathbf{W}_z \mathbf{p}_z = \lambda_z^2 \;,
\end{equation}
\begin{equation}\label{observation_2}
\sum_{i,j} (p_i p_j \delta_{z_i z}\delta_{z_j z})^2 = \lambda_z^2  \;,
\end{equation}
the aforementioned objective function (Eq. (\ref{GSBM-P-OQ})) can be expressed as follows:
\begin{equation}\label{OF_Model2}
Q_\mathrm{GSBM-P} = \sum_z \lambda_z^2 \;.
\end{equation}

Therefore, the maximum likelihood estimation of GSBM-P is equivalent to maximizing the sum of the squares of the principal eigenvalues of the adjacency matrices of all communities. Intuitively, a community $z$ with higher $\lambda_z$ is better intra-connected. Thus, the maximum likelihood estimation of GSBM-P aims to find communities that are densely intra-connected.

\subsubsection{Relationship with Label Propagation Algorithm with node preference}

Rewriting the objective function according to Eq. (\ref{observation_1}) and Eq. (\ref{observation_2}) yields
\begin{equation}\label{OF_Model2_alter}
Q_\mathrm{GSBM-P} = \sum_z \sum_{i,j} p_i p_j W_{ij} \delta_{z_i z} \delta_{z_j z}  \ ,
\end{equation}
which is the same as the objective function of general label propagation with node preference (i.e., Eq. (\ref{Q_LPA-P})).

We describe the adopted coordinate ascent method here in detail. This method optimizes the objective function in a similar way to label propagation algorithm. Coordinate ascent method aims to maximize an objective function $f(X)$, where $X$ is the collection of all parameters. Coordinate ascent method adjusts the value of one parameter to maximize objective function while fixed all other parameters. Different parameters are adjusted cyclically. It is guaranteed that objective function is non-decreasing during this process and the method converges to a local optimum where objective function no longer increases.

To optimize the objective function of our model, our coordinate ascent method treats both the block indicators (i.e., labels of vertices) and node preference $\mathbf{p}$ as parameters. To adjust node preference, we prefer to adjust the node preference $\mathbf{p}_z$ of all vertices inside one community $z$, by approximating the product of $\sqrt{\lambda_z}$ and the $L_2$-normalized local eigenvector centrality computed inside the community $z$. To adjust label of one vertex given all other vertices' labels and $\mathbf{p}$, the coordinate ascent method updates label of vertex $i$ as follows:
\begin{equation}\label{LPPA_update}
z_i = \arg\max_z \sum_{j \neq i} p_j W_{ij} \delta_{z_j z} \ ,
\end{equation}
which is exactly the update rule of label propagation with node preference without considering the hop attenuation regularization.
Note that assigning a new label for vertex $i$ will not increase the objective function given all other vertices' labels and $\mathbf{p}$.
The difference between our method and general label propagation with node preference is that the node preference in Eq. (\ref{p_update}) is derived in a statistically-grounded way rather than heuristically defined.
The update rule has a clear explanation. Each vertex has a similarity with each community, which is the weighted sum of linked edge weights connected to this community, with more representative (higher local eigenvector centrality) vertex in the community owning higher weight. Then, each vertex joins in the most similar community. In this sense, the node preferences of vertices inside two communities should be updated immediately when a vertex joins in a new community, though it is not required by the coordinate ascent method.

The objective function in Eq. (\ref{OF_Model2_alter}) is non-decreasing during the process of estimating labels and node preference of vertices. The coordinate ascent method is terminated when no vertex changes label or the objective function no longer increases after several iterations. The number of communities is determined as the number of different labels in the converged state.

\subsection{A Constrained Version of GSBM-P}\label{Model1}

In this section, we examine a constrained version of our model and show its relation to label propagation under constraint. The constrained version assumes that each vertex $i$ inside the community $z$ owns the same node preference $\sqrt {\mu_z}$, and hence is not degree corrected. That is,
\begin{equation}
W_{ij} \sim \mathcal{N}(\mu_{z_i} \delta_{z_i z_j}, \sigma^2) \;,
\end{equation}
where the expected edge weight between vertex $i$ and $j$ is $\mu_{z_i}\delta_{z_i z_j}$ or equivalently
$\sqrt {\mu_{z_i}}\sqrt {\mu_{z_j}}\delta_{z_i z_j}$. $\mu_{z_i}$ is a parameter associated with each community.
This simplified model assumes that the expected value of the weights of the inter-community edges is 0;
the expected value of the weights of the edges inside community $z$ is a uniform value $\mu_z$.

The likelihood function of a graph under this constrained model is
\begin{equation}
P(W|\mathbf{z}, \bm{\mu} ,\sigma ) = \prod\limits_{i,j} {\frac{1}{{\sigma \sqrt {2\pi } }}} \exp \frac{{ - {{\left( {{W_{ij}} - {\mu _{{z_i}}}\delta_{z_i z_j}} \right)}^2}}}{{2{\sigma ^2}}}\;.
\end{equation}
% \boldsymbol{\mu}

Maximizing the likelihood function is equivalent to minimizing $\sigma$ for Gaussian models.
The maximum likelihood estimation of $\sigma^2$ can be expressed as
\begin{equation}
\begin{array}{rcl}
\sigma^2 &=& \displaystyle \frac{\sum_{i,j}(W_{ij} - \mu_{z_i}\delta_{z_i z_j})^2}{n^2} \\
&=& \displaystyle \frac{-2 \sum_{i,j} \mu_{z_i} W_{ij} \delta_{z_i z_j}  + \sum_{i,j}  (\mu_{z_i}\delta_{z_i z_j})^2 }{n^2} + \mathrm{const} \;.
\end{array}
\end{equation}
Hence, minimizing $\sigma^2$ is equivalent to maximizing the following objective function:
\begin{equation}\label{M1Q-original}
\begin{array}{rcl}
Q &=& \displaystyle 2\sum_{i,j} \mu_{z_i} W_{ij}\delta_{z_i z_j} - \sum_{i,j}  (\mu_{z_i}\delta_{z_i z_j})^2 \\
  &=& \displaystyle 2 \sum_z \sum_{i,j} \mu_z W_{ij} \delta_{z_i z} \delta_{z_j z} -\sum_z \mu_z^2 n_z^2 \ ,
\end{array}
\end{equation}
where $n_z$ denotes the number of vertices inside community $z$.

In this constrained version of our model, if we specify $\mu_z=\mu$ for all communities, then this simplified quality function can be viewed as the objective function of label propagation algorithm under constraint, since
\begin{equation}
\begin{array}{rcl}
Q &=& \displaystyle \mu \left( 2 \sum_z \sum_{i,j}  W_{ij} \delta_{z_i z} \delta_{z_j z} - \mu \sum_z n_z^2 \right) \\
  &=& \displaystyle \mu \left( 2 Q_{\mathrm{LPA}} - \mu \sum_z n_z^2 \right) \ ,
\end{array}
\end{equation}
where $\sum_z n_z^2$  is the penalty term, which is maximized when all vertices are in a community, and prevents the large communities from growing. $\mu$ can be viewed as a resolution parameter, which controls the community size.

In comparison with the first model of label propagation algorithm under constraint \cite{barber2009detecting} (a.k.a. constant Potts model \cite{traag2011narrow}), leaving $\mu$ tunable makes two objective functions equivalent.
Therefore, the maximum likelihood estimation of the block indicators in the constrained version of our model while leaving $\mu_z$ tunable results to an objective function which is a generalization of the objective function of label propagation algorithm proposed by Barber and Clark \cite{barber2009detecting} with a penalty term $\frac{1}{2} \sum_z n_z^2$.

The constrained version of our model has its disadvantages, especially that the expected weight of edges inside each community is equal. It may not fit well in real networks. By incorporating node preference, the expected weight of edges may vary according the the node preference assigned for each vertex. This renders our model more expressive and robust against modelling complex real-world networks. We show in next section that our proposed Gaussian stochastic blockmodel with node preference performs well.

\section{Experiments}\label{SectionExperiments}

In this section, the coordinate ascent method for GSBM-P is tested on both real-world networks and synthetic networks. We also compare it with label propagation algorithm (LPA), label propagation algorithm with intra-community RandomWalk (LPA-P) as the node preference and OSLOM \cite{lancichinetti2011finding} on various synthetic networks, and the variational Bayes method for weight stochastic blockmodel (WSBM) on weighted synthetic networks.

Two real-world networks, namely the \textit{karate club} \cite{zachary1977information} and \textit{political blogs} \cite{adamic2005political} are chosen as test data. For synthetic networks, unweighted benchmark networks proposed by Lancichinetti et al. \cite{lancichinetti2008benchmark} and weighted benchmark networks proposed by Lancichinetti and Fortunato \cite{lancichinetti2009benchmarks} are chosen. Erd\"{o}s-R\'enyi random graphs \cite{erdos1959random} are chosen for checking overfitting. Gaussian distribution is not suitable to fit binary data (unweighted networks). However, in the literature, label propagation algorithms are often applied in unweighted networks. Due to the same objective function as label propagation with node preference and similar optimization method to label propagation with node preference, we also test our coordinate ascent method for GSBM-P in unweighted networks and show its good performance.

Normalized mutual information (NMI, a.k.a. symmetric uncertainty, introduced by Witten and Frank \cite{witten2005data}) is often adopted to reflect the similarity between the obtained partition and the planted partition. However, as is shown in \cite{zhang2015revisit}, NMI sometimes has systematic bias in finite-size networks. This phenomena is also observed in our experiments on both synthetic unweighted networks with no intra-community edge in planted partitions and synthetic weighted networks with total intra-community edge weights close to zero in planted partitions. Algorithms that are designed only to find densely intra-connected communities can not find similar partitions to the planted partitions on these networks. However, for LPA-P and the coordinate ascent method for GSBM-P, the values of NMI are much larger than 0 between obtained partitions and the planted partitions on these networks (the presentation of results is omitted).

Zhang \cite{zhang2015revisit} proposed relative normalized mutual information (rNMI) to fix this systematic bias. rNMI is defined as follows:
\begin{equation}
rNMI\left(A,B\right)=NMI\left(A,B\right)-\langle NMI\left(A,C\right) \rangle \ ,
\end{equation}
where $A$ is the planted partition, $B$ is partition obtained by algorithm, $C$ is a random partition with the same group-size distribution with partition $B$, and $\langle NMI\left(A,C\right) \rangle$ is the expected $NMI\left(A,C\right)$ over different realizations of random partition $C$. In our experiments, we observe that the value of rNMI is slightly smaller than zero for LPA-P and the coordinate ascent method for GSBM-P on these benchmark networks with no intra-community edge or with total intra-community edge weights close to zero in planted partitions.

For synthetic networks, to reflect the similarity between the obtained partition $B$ and the planted partition $A$, we will use the ratio of relative normalized mutual information (rrNMI) as defined below:
\begin{equation}\label{rrNMI}
rrNMI\left(A,B\right)=\frac{rNMI\left(A,B\right)}{rNMI\left(A,A\right)} ,
\end{equation}
such that it is up-bounded by 1 and equals 1 when partition $B$ is identical to partition $A$. Note that $rNMI\left(A,A\right)$ is not zero for all the planted partitions in our experiment, such that this equation will not suffer from divide-by-zero problem. The expectation in rNMI is estimated over 100 realizations of random partition $C$.

OSLOM \cite{lancichinetti2011finding} may produce overlapping communities. Both NMI and rrNMI can not be directly applied. An adjusted NMI \cite{lancichinetti2009detecting} is used to measure the similarity between the obtained covering by OSLOM and the planted partition. The adjusted NMI equals 1 when obtained covering and planted partition are identical.

In all of the following experiments, LPA-P is terminated after 50 iterations. The coordinate ascent method for GSBM-P starts from the initialization where each vertex has its unique label. The iterative order of the vertices for the coordinate ascent method is random, and the node preferences inside two communities are updated immediately when a vertex moves from one community to another. Our method is terminated when no vertex changes label or the objective function no longer increases. We observe that the actual number of iterations of our method is generally less than 10 in tested networks. The time complexity of our method should be $\mathcal{O}((d+td_1C)n)$ in each iteration, where $d$ denotes the average degree, $t$ denotes the average iterative times for computing intra-community eigenvector centrality, $d_1$ denotes the average intra-community degree, and $C$ denotes the average size of communities.
In very large networks, the node preferences can be updated everytime all vertices are traversed for saving time. Then, the time complexity should be $\mathcal{O}(dn+td_1n)$ in each iteration, which is scalable for very large networks.

\subsection{Empirical networks}

\textit{Karate club} data set \cite{zachary1977information} is a social network of friendships between 34 members of a karate club at a U.S. university in the 1970s. The network has 34 nodes and 78 edges with weight 1. In reality, the karate club finally splits into 2 groups. The true partition of this network is known.

The coordinate ascent method for GSBM-P is run several times and the partition with the highest quality function value is chosen.

\begin{figure}
  \centering
  \includegraphics[width=3in]{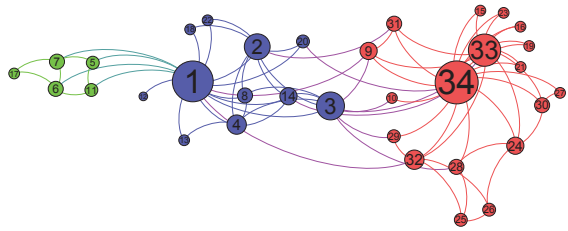}\\
  \caption{Partitions of karate club network by GSBM-P. The size of each vertex is proportional to its degree and the color reflects the group membership.}\label{karate_model2}
\end{figure}

Fig. \ref{karate_model2} demonstrates the result of our method in the karate club network.

The result of our method is a more fine-grained partition of the true partition. In fact, this result has a slightly higher quality function value than the true partition. Note that modularity maximization method in \cite{agarwal2008modularity} and the non-parametric Bayesian mixed membership stochastic blockmodel \cite{koutsourelakis2008finding} favor the partition with four communities. The true partition is a local optimum of our method which is frequently reached.

We notice that the vertex 10 is misidentified by degree corrected stochastic blockmodel in \cite{karrer2011stochastic} and modularity maximization method in \cite{agarwal2008modularity}. We verify that the objective function value (i.e., Eq. (\ref{OF_Model2}) or Eq. (\ref{OF_Model2_alter})) of the true partition is higher than that of the partition with misidentification of vertex 10. Thus GSBM-P performs better in karate club network.

For GSBM-P, we observed that vertices with higher degree have generally higher node preference inside one community, and in karate club network, the pair of vertices with higher degree inside one community are more likely to be connected. It is the same as expected by the model, which expects the weight of edge linked to pair of vertices with higher node preference should be larger.

\begin{figure}
  \centering
  \includegraphics[width=3in]{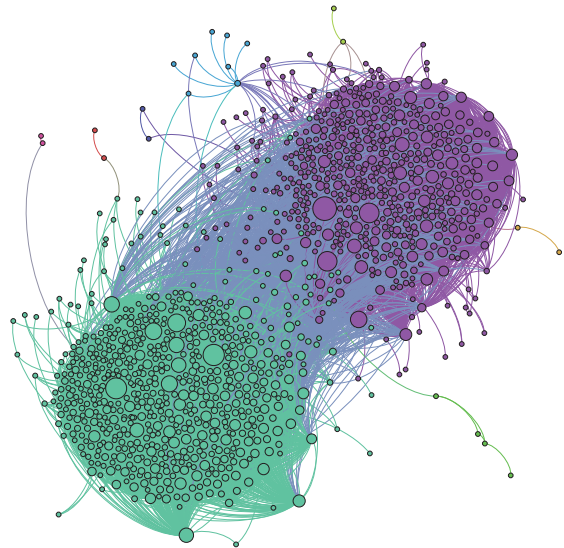}\\
  \caption{Partition of \textit{political blogs} network by GSBM-P. The size of each vertex is proportional to its degree and the color reflects the group membership.}\label{polblogs_model2}
\end{figure}

We then show how our method performs in a larger network.
\textit{political blogs} data set \cite{adamic2005political} is a network of directed hyperlinks between political blogs whose largest connected component contains 1222 nodes. In this paper, we use the undirected form.
Fig. \ref{polblogs_model2} shows the result of our method in the larget connected component of political blog network.

GSBM-P discovers 9 communities with 7 tiny communities and 2 major communities that are roughly two political tendencies.
The normalized mutual information of obtained partition is 0.678.
If we merge small communities to one of the two major communities, normalized mutual information increases to 0.724. The performance is very close to that of the degree corrected stochastic blockmodel  \cite{karrer2011stochastic} with given cluster number.

\subsection{Synthetic networks}

\subsubsection{Unweighted synthetic networks}

\begin{figure}[htbp]
\centering
\subfigure[]{
\begin{minipage}[t]{0.45\linewidth}
\centering
\includegraphics[width=3in]{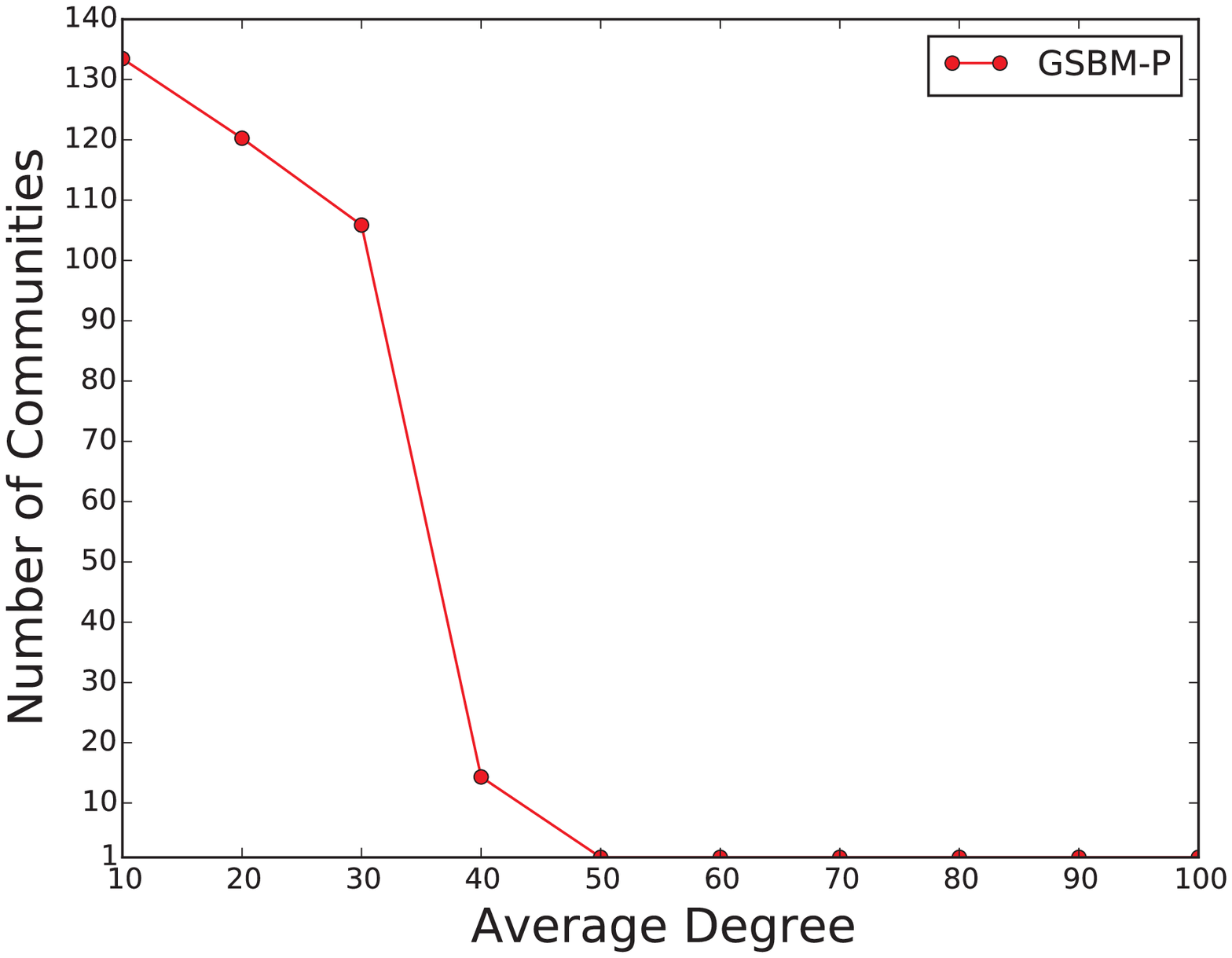}
\label{ERgraphdchange}
\end{minipage}%
}
\subfigure[]{
\begin{minipage}[t]{0.45\linewidth}
\centering
\includegraphics[width=3in]{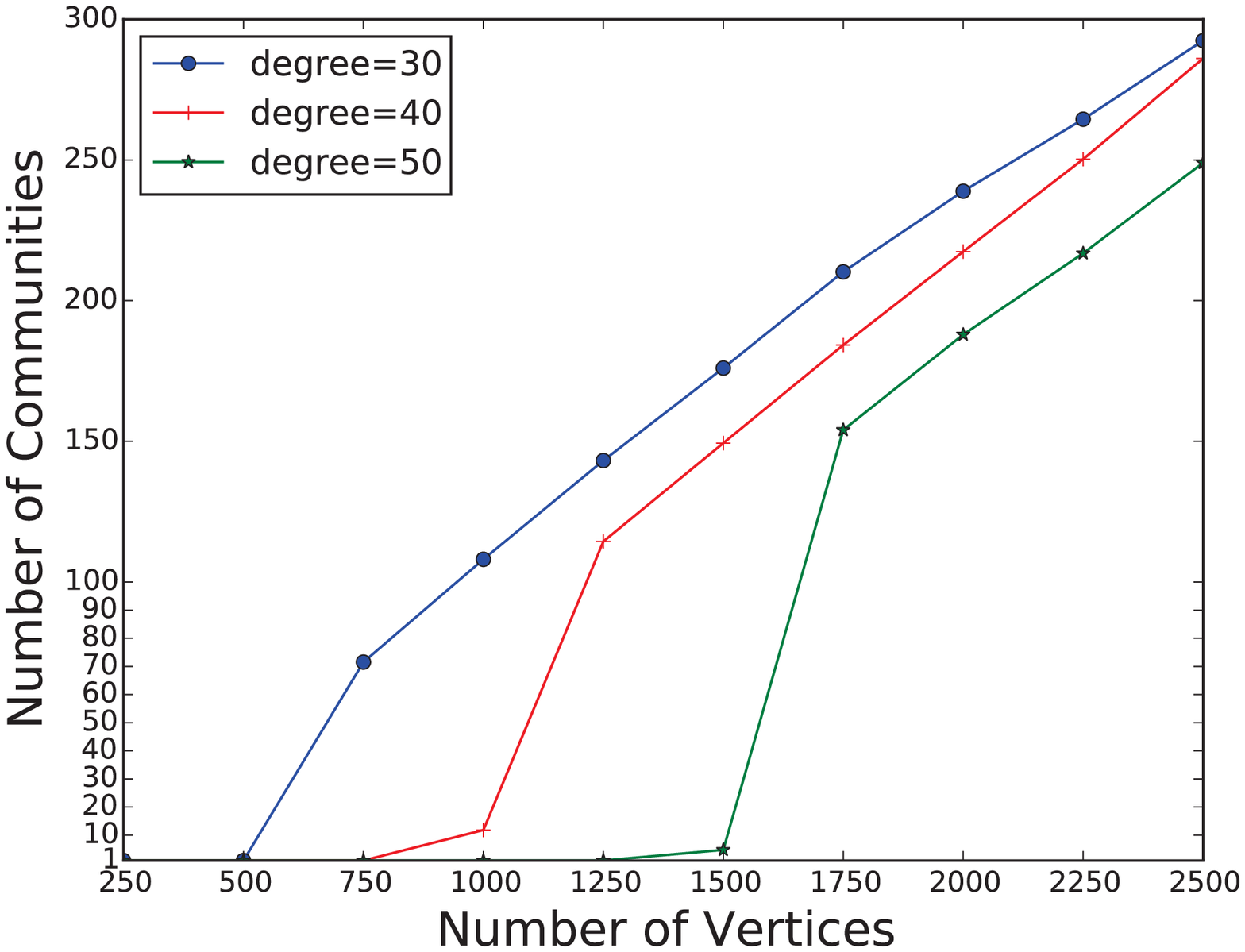}
\label{ERgraphnchange}
\end{minipage}
}
\caption{Overfitting checking on Erd\"{o}s-R\'enyi random graphs \cite{erdos1959random}. Each data point is an average over 30 graphs. In (a), we show the discovered number of communities with respect to the average degree. The number of vertices in each graph is fixed to 1000. In (b), we show the discovered number of communities with respect to the system size. }
\label{ER}
\end{figure}

Since our model treats the block indicators as parameters, the maximum likelihood estimation of the block indicators of vertices may be bias especially when these vertices have very small degree.
To check overfitting, the coordinate ascent method for GSBM-P is tested on Erd\"{o}s-R\'enyi random graphs \cite{erdos1959random} where no community structures exist. Analogous to \cite{lancichinetti2009community}, the network sizes are fixed to 1000 nodes, and the average degrees in all random graphs range from 10 to 100. Our method is run 10 times on each random graph and the results with highest objective function value are chosen. The number of discovered communities in each random graph is illustrated in Fig. \ref{ERgraphdchange}. It shows that the coordinate ascent method is not overfitting when the average degree is not too small. In Erd\"{o}s-R\'enyi random graphs with network size equal to 1000, our method begins to identify the whole network as a community when the average degree is 40, though it fails in some generated Erd\"{o}s-R\'enyi random graphs with average degree equal to 40.

We also show how number of communities found by our method varies with respect to the network size of Erd\"{o}s-R\'enyi random graphs in Fig .\ref{ERgraphnchange}. It can be observed that when the system size of Erd\"{o}s-R\'enyi random graphs increases, our method demands larger average degree to find only one community in Erd\"{o}s-R\'enyi random graphs. It implies that our method may overfit in networks with large communities when the intra-community degree of vertices remains the same, in that it may not recognize planted community as only one community. To deal with this situation, one can evaluate the significance of obtained communities and merge the less significant communities into other communities. This process will decrease the number of communities. One property of our method and label propagation algorithms is that the number of communities will not increase in the iterations. Thus, it is reasonable to apply our method again after the mergence procedure.

\begin{figure}
  \centering
  \includegraphics[width=3in]{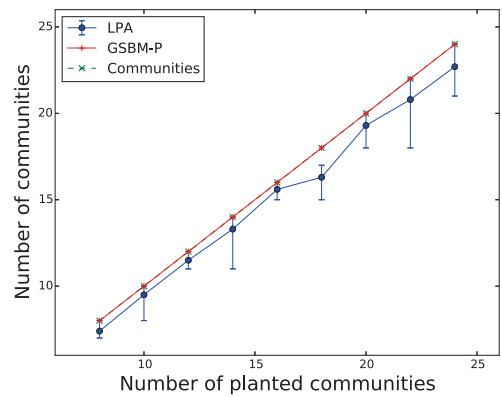}\\
  \caption{Resolution limit test on benchmark networks \cite{fortunato2007resolution}. The networks are composed by cliques with 4 vertices and each clique is linked to the next clique with an edge to form a ring. We show the number of discovered communities in networks composed by different number of cliques varying from 8 to 24. Each data point is an average over 10 runs.}
\label{resolution}
\end{figure}

Our method is also tested on resolution limit test benchmark networks \cite{fortunato2007resolution}. Analogous to \cite{vsubelj2011generalized}, the networks are composed by cliques with 4 vertices and each clique is linked to the next clique with an edge to form a ring. Both of our method and label propagation algorithm \cite{raghavan2007near} are run 10 times on each network and the average number of discovered communities is shown in Fig. \ref{resolution}. Each time, our method discovers the correct communities. Hence, in Fig. \ref{resolution}, the red line covers the green dashed line which represents the number of planted communities. Thus, our method tends not to suffer from resolution limit. However, the blue error bars in Fig. \ref{resolution} imply that LPA algorithm is not stable. The average number of discovered communities of LPA is less than the planted, because LPA sometimes identifies several cliques as one community.

We then test the algorithms in the unweighted benchmark networks proposed by Lancichinetti et al. \cite{lancichinetti2008benchmark}. Experiments show that the coordinate ascent method for GSBM-P is superior to LPA in those networks, and achieves a similar performance to OSLOM.

There are several parameters to generate the benchmark networks, including the number of vertices $n$, average degree, maximum degree, the degree distribution, the community size distribution, the range of community size $C$, and the ratio that a vertex links to vertices outside its community (i.e., topological mixing parameter $\mu_t$). When the mixing parameter is smaller than 0.5, the communities in generated networks are strong communities \cite{radicchi2004defining} where each vertex has more connections with vertices inside its community than vertices outside the community. When topological mixing parameter $\mu_t=1$ in benchmark networks, communities of planted partition have no intra-community edges.

\begin{figure}[htbp]
\centering
\subfigure[]{
\begin{minipage}[t]{0.45\linewidth}
\centering
\includegraphics[width=3in]{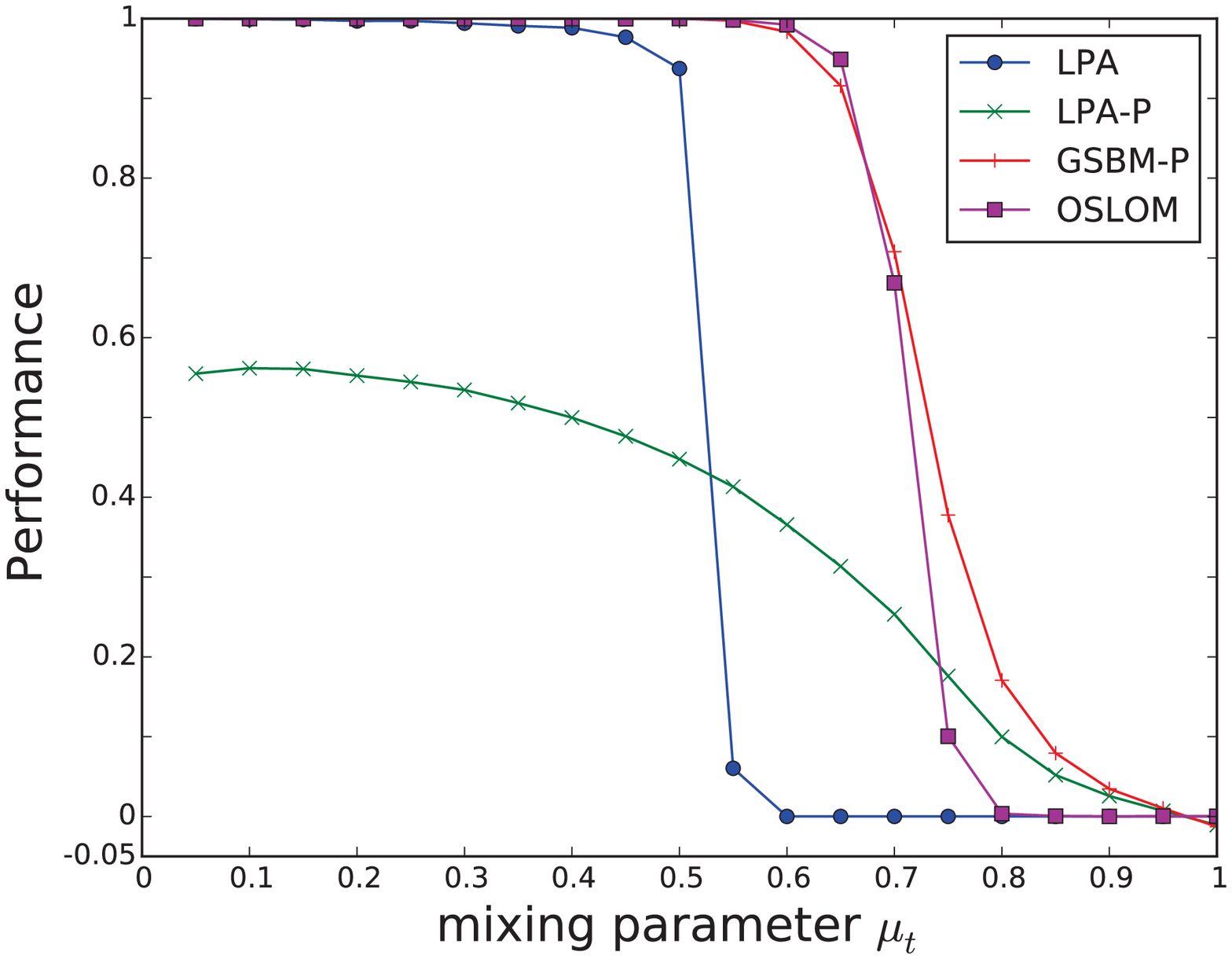}
\label{figure_1}
\end{minipage}%
}
\subfigure[]{
\begin{minipage}[t]{0.45\linewidth}
\centering
\includegraphics[width=3in]{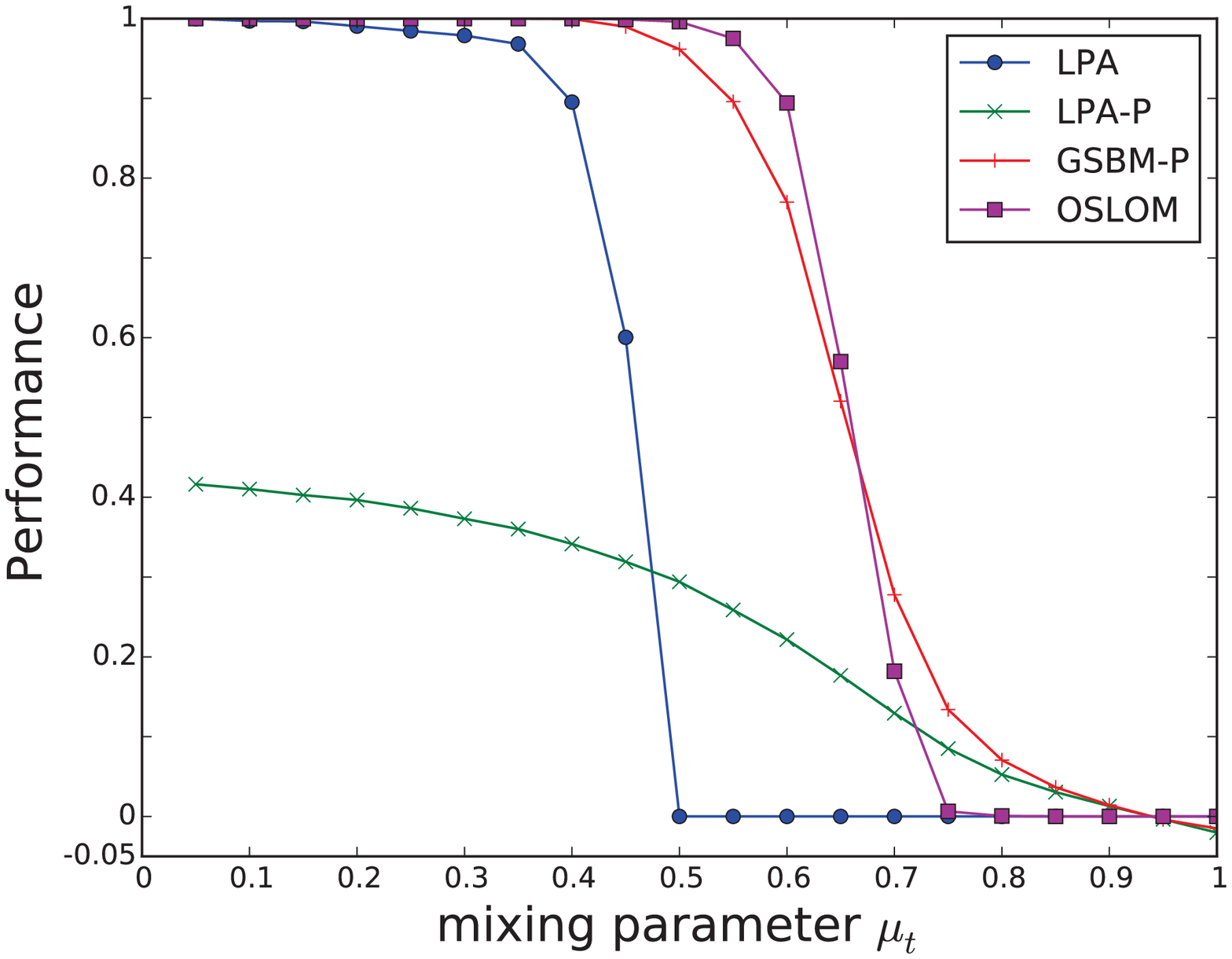}
\label{figure_2}
\end{minipage}
}
\caption{Performance with respect to the mixing parameter $\mu_t$ for different algorithms on two unweighted test benchmarks \cite{lancichinetti2008benchmark}. We use ratio of relative normalized mutual information (Eq. (\ref{rrNMI})) for LPA, LPA-P and GSBM-P. We use adjusted NMI \cite{lancichinetti2009detecting} for OSLOM. Each data point is an average over 30 networks generated by the same parameters. In two subfigures, the size of networks is 1000, the average degree is 20, and the maximum degree is 50. In (a), the range of community size is $C=[10,50]$. In (b), the range of community size is $C=[20,100]$.}
\label{figure_unweighted}
\end{figure}

We set two kind of parameters with different range of community size according to \cite{lancichinetti2009community} while leaving the mixing parameter $\mu_t$ as the independent variable. We compare GSBM-P with LPA-P and LPA. The ratio of relative normalized mutual information (Eq. (\ref{rrNMI})) between the obtained partition and planted partition is calculated. For OSLOM, adjusted NMI \cite{lancichinetti2009detecting} is calculated. Fig. \ref{figure_unweighted} shows how the performance of these algorithms vary according to the mixing parameter in two kind of networks.

From Fig. \ref{figure_1} and Fig. \ref{figure_2}, it can be seen that the planted partition is an optimum of our method when the mixing parameter $\mu_t$ is small. When mixing parameter $\mu_t$ is small, coordinate ascent method for GSBM-P always converges to the optimum in every run. Our method outperforms LPA. The label propagation approach applied in LPA is very suitable for finding strong communities. However, as the mixing parameter is larger than 0.5 (i.e., weak communities), LPA fails.
The performance of LPA-P is not satisfying.

We can not directly compare the performance of our coordinate ascent method for GSBM-P with OSLOM since their performance are evaluated by different measures. However, values of both measures equal one when the obtained partition or covering is identical to planted partition. Fig. \ref{figure_1} shows that our coordinate ascent method recovers planted partition in a similar range of topological mixing parameters to OSLOM in networks with small communities. Fig. \ref{figure_2} shows that the performance of our coordinate ascent method is close to that of OSLOM in networks with large communities.

\begin{figure}[htbp]
\centering
\subfigure[]{
\begin{minipage}[t]{0.45\linewidth}
\centering
\includegraphics[width=3in]{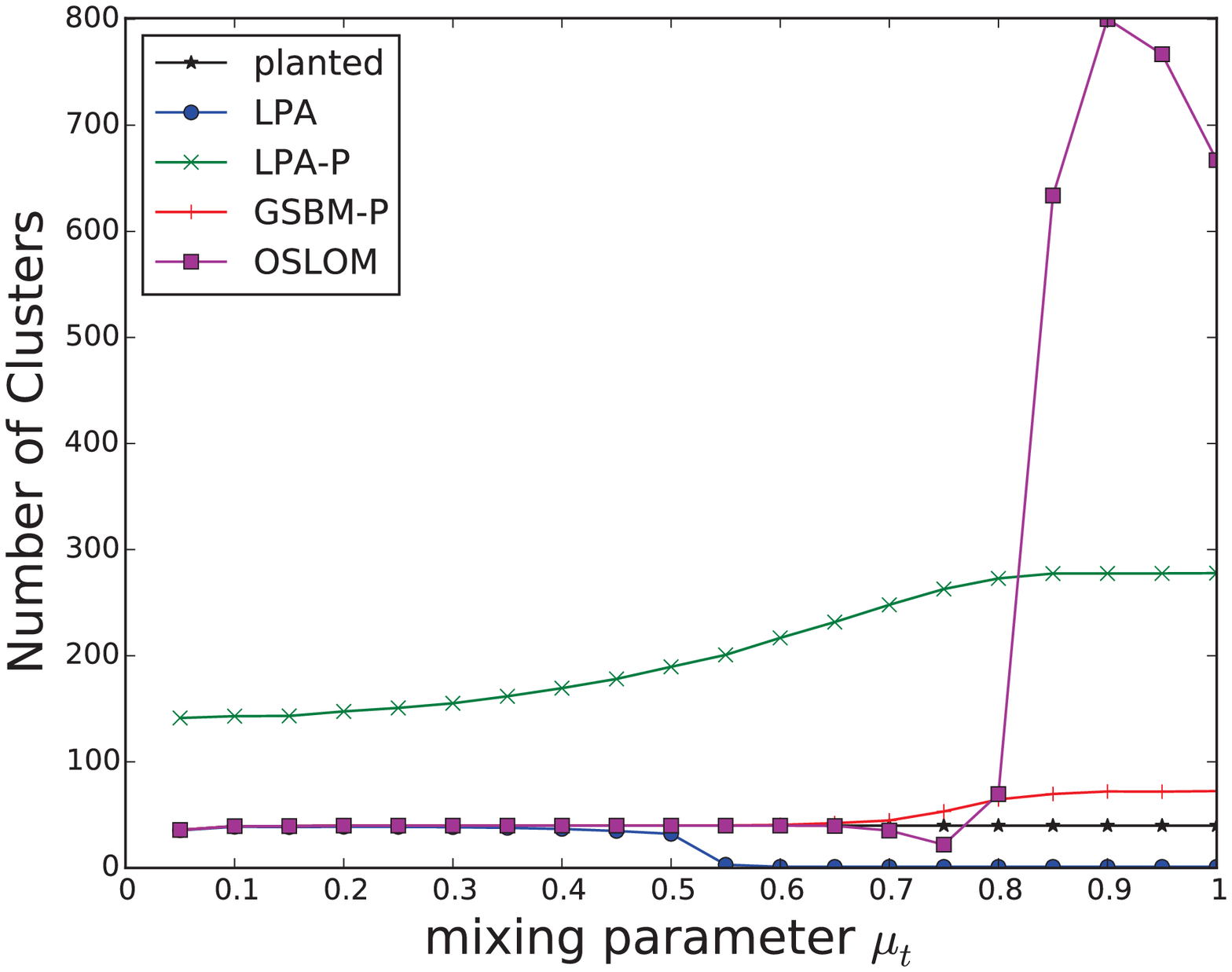}
\end{minipage}%
}
\subfigure[]{
\begin{minipage}[t]{0.45\linewidth}
\centering
\includegraphics[width=3in]{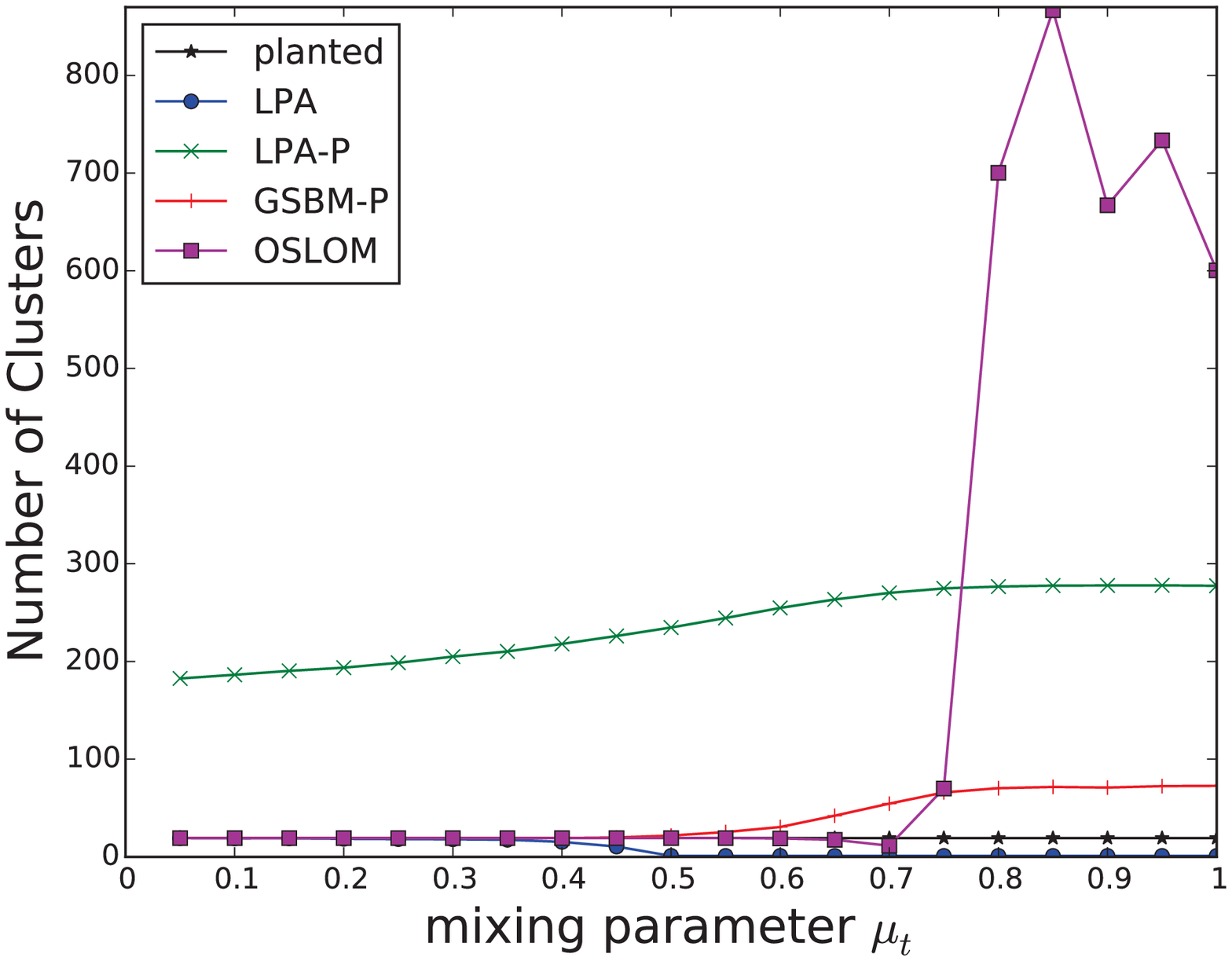}
\end{minipage}
}
\caption{Number of communities found with respect to the mixing parameter $\mu_t$ for different algorithms on two unweighted test benchmarks \cite{lancichinetti2008benchmark}. Each data point is an average over 30 networks generated by the same parameters. In two subfigures, the size of networks is 1000, the average degree is 20, and the maximum degree is 50. In (a), the range of community size is $C=[10,50]$. In (b), the range of community size is $C=[20,100]$.}
\label{unweighted_clusternum}
\end{figure}

We then show the corresponding average number of communities found by different algorithms in Fig. \ref{unweighted_clusternum}. The average number of planted communities is also shown. LPA discovers one community when topological mixing parameter $\mu_t$ equals or is larger than 0.6 in benchmark networks with small communities, and equals or is larger than 0.5 in benchmark networks with large communities. In benchmark networks with small communities, our coordinate ascent method for GSBM-P discovers reasonable number of communities in a similar range of topological mixing parameters to OSLOM. However, in benchmark networks with large communities, our method discovers more communities when $\mu_t$ equals or is larger than 0.55. As discussed above, it may improve the performance of our method if mergence procedure is introduced.

\begin{figure}[htbp]
\centering
\subfigure[]{
\begin{minipage}[t]{0.45\linewidth}
\centering
\includegraphics[width=3in]{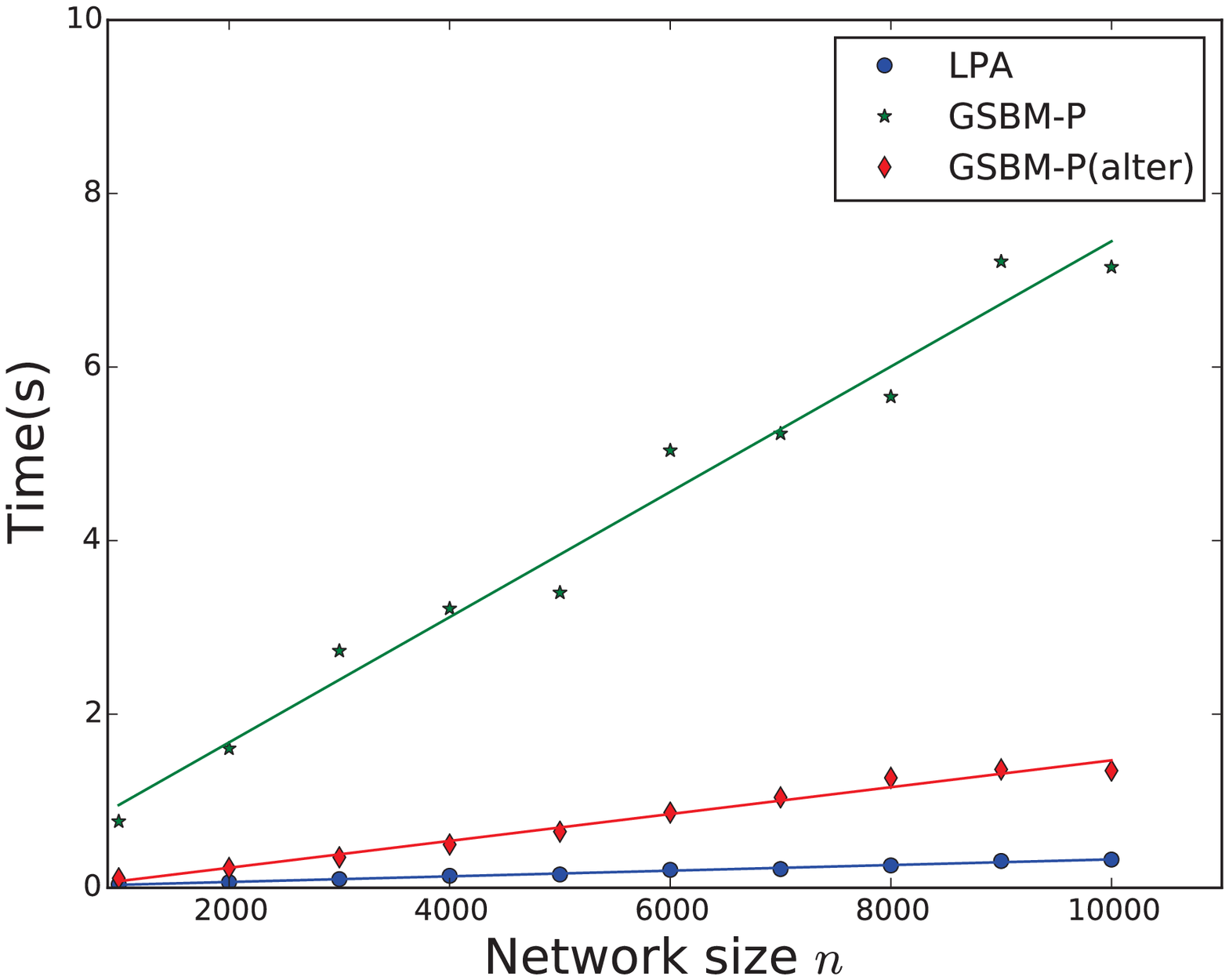}
\label{figure_ntime}
\end{minipage}%
}
\subfigure[]{
\begin{minipage}[t]{0.45\linewidth}
\centering
\includegraphics[width=3in]{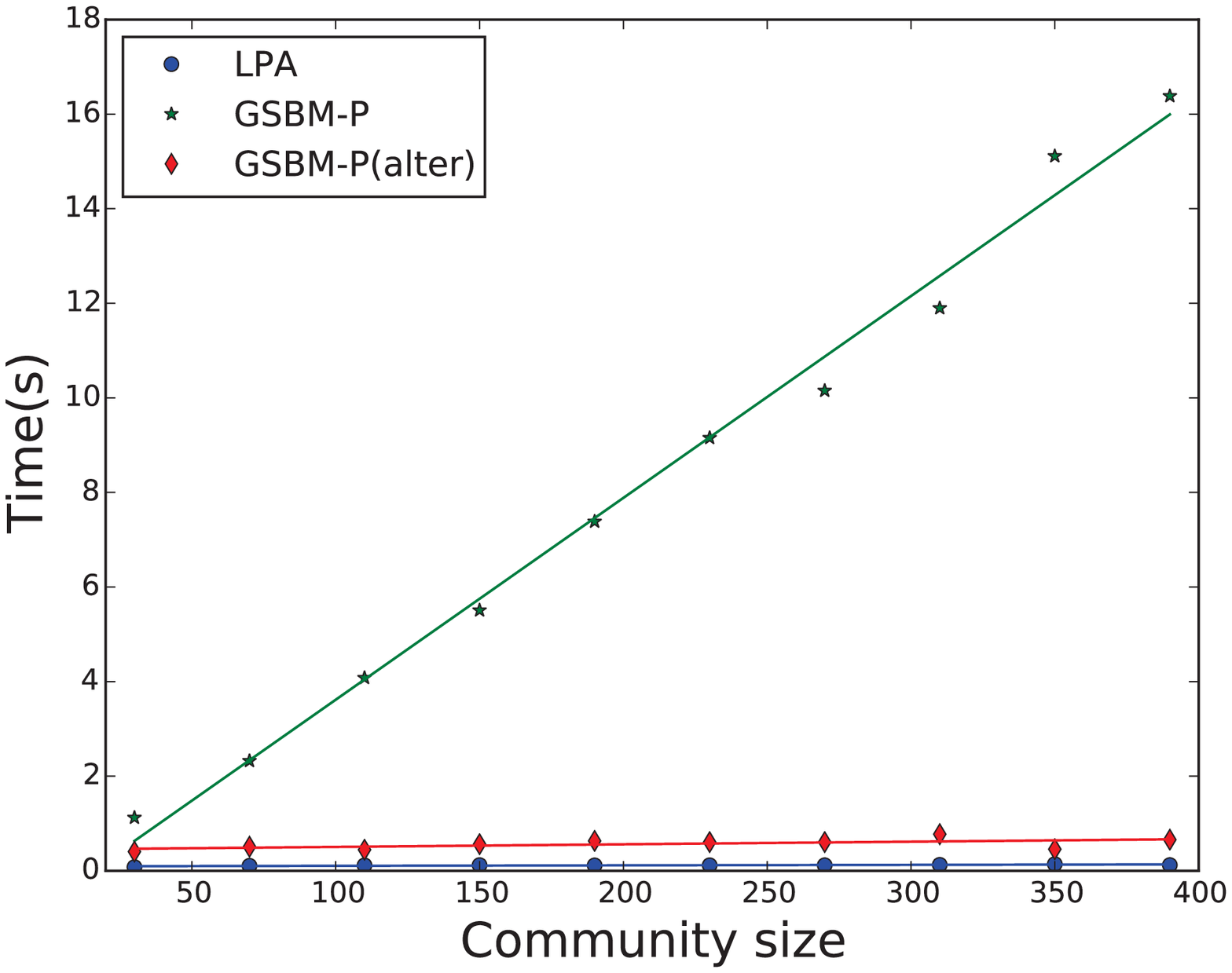}
\label{figure_ctime}
\end{minipage}
}
\caption{Running time with respect to the network size and community size for different algorithms on unweighted test benchmarks \cite{lancichinetti2008benchmark}. Each data point is an average over 100 runs. In two subfigures, the average degree is 20 ,the maximum degree is 50 and the topological mixing parameter $\mu_t$ is 0.1. In (a), the range of community size is $C=[20,100]$. In (b), the size of networks is 3000.}
\label{figure_time}
\end{figure}

We then show the running time of algorithms in Fig. \ref{figure_time}. In Fig. \ref{figure_time}, GSBM-P(alter) represents the alternative coordinate ascent method for GSBM-P that only updates node preferences everytime all vertices are traversed. Mixing parameter $\mu_t$ is set to 0.1, while parameters other than the number of vertices and community size are setting according to \cite{lancichinetti2009community}. In Fig. \ref{figure_ntime}, the community sizes are fixed to range from 20 to 100, and the number of vertices varies among the test networks. In Fig. \ref{figure_ctime}, network sizes are fixed to 3000, the community size varies. The community size in $x$ axis represents the average value, for instance, 390 in $x$ axis represents the community sizes ranges from 370 to 410, and so forth.

It shows that the running time of our method is linear to network size and community size. For very large networks with large communities, the alternative coordinate ascent method can be applied.

\subsubsection{Weighted synthetic networks}

We then test the algorithms in the weighted benchmark networks proposed by Lancichinetti and Fortunato \cite{lancichinetti2009benchmarks}. The weights of edges are non-negative in these benchmark networks.

There are two more parameters in weighted benchmark networks than unweighted benchmark networks, namely $\beta$ which controls the power-law relation between the sum of weights a vertex links to its neighbors and the vertex's degree, and the ratio of weights a vertex links to vertices outside its community (i.e., mixing parameter $\mu_w$).

\begin{figure*}[!htbp]
\centering
\subfigure[]{
\begin{minipage}[t]{0.45\linewidth}
\centering
\includegraphics[width=3in]{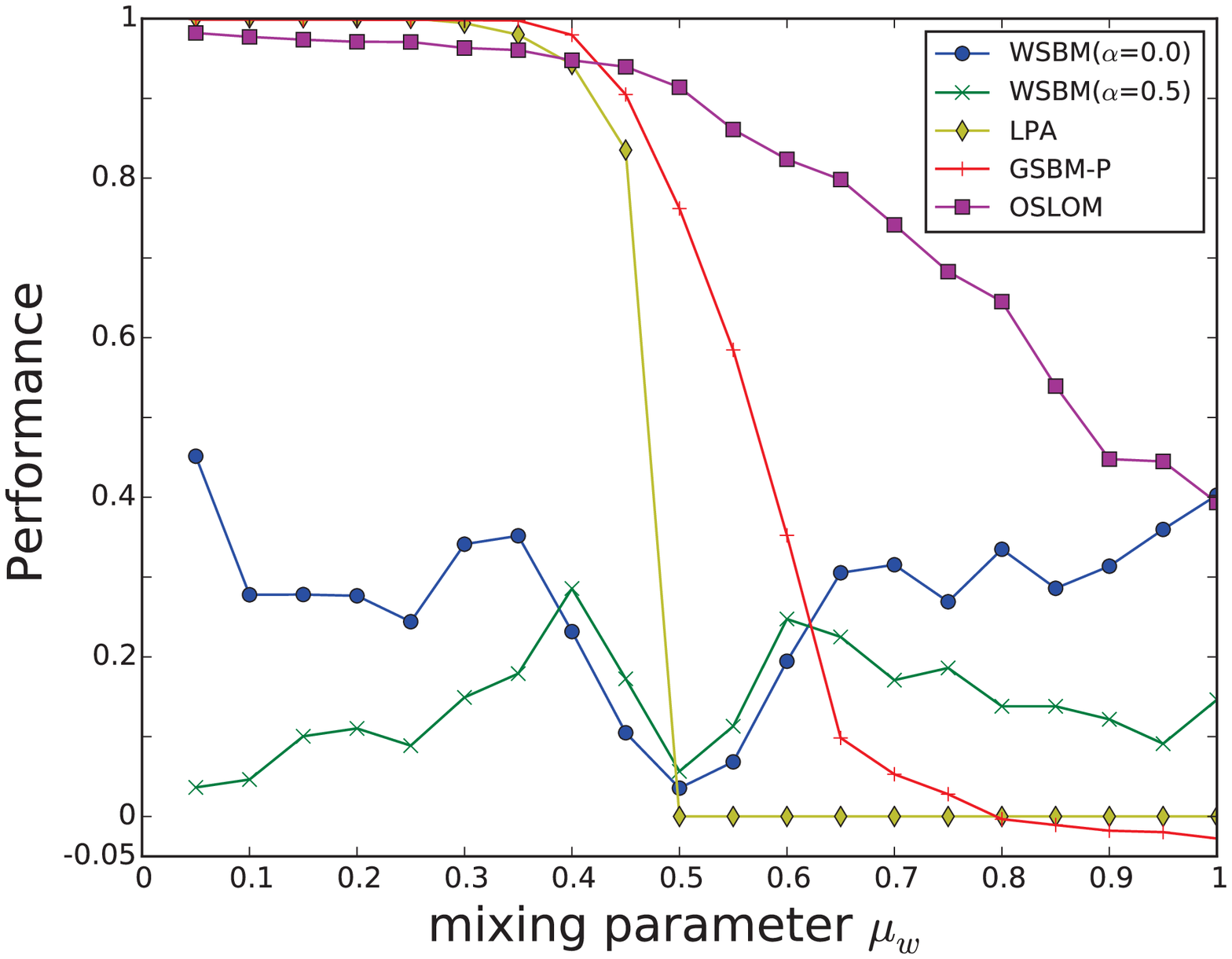}
\label{figure_7}
\end{minipage}
}
\subfigure[]{
\begin{minipage}[t]{0.45\linewidth}
\centering
\includegraphics[width=3in]{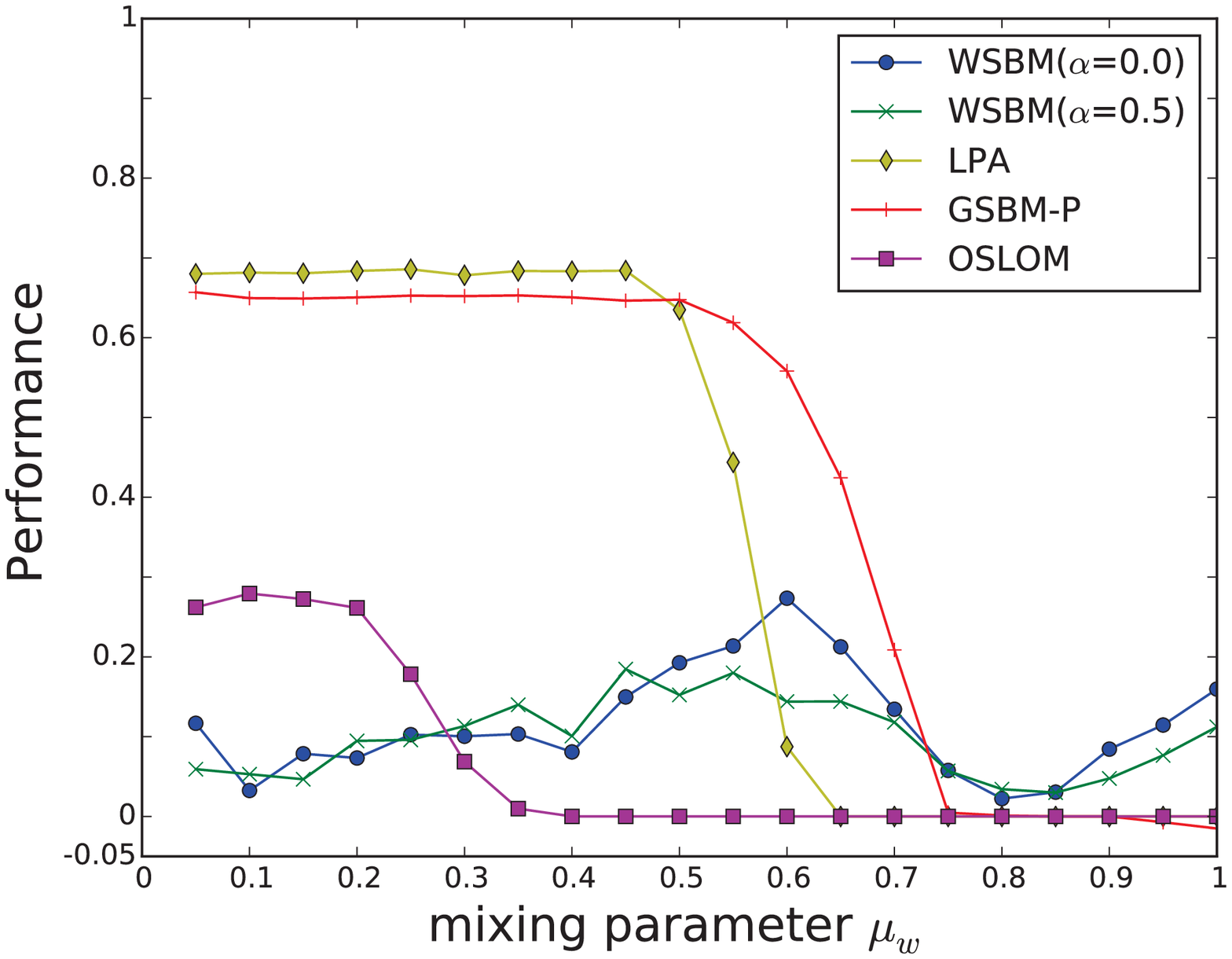}
\label{figure_8}
\end{minipage}
}
\caption{Performance with respect to the mixing parameter $\mu_w$ for different algorithms on two small weighted test benchmarks \cite{lancichinetti2009benchmarks}. We use ratio of relative normalized mutual information (Eq. (\ref{rrNMI})) for WSBM, LPA and GSBM-P. We use adjusted NMI \cite{lancichinetti2009detecting} for OSLOM. Each data point is an average over 30 networks generated by same parameters.  In (a) and (b), the size of networks is 150, the average degree is 15, the maximum degree is 30, and the range of community size is $C=[10,20]$. In (a), the topological mixing parameter $\mu_t$ is 0.5. In (b), the topological mixing parameter $\mu_t$ is 0.8.}
\label{figure_small_weighted}
\end{figure*}

We compare our method with LPA and the variational Bayes method for WSBM. We set the parameter $\alpha$ in WSBM (see Eq. (\ref{WSBM})) to 0.0 and 0.5, corresponding to pure WSBM and WSBM with auxiliary degree correction respectively. Number of planted communities is given for WSBM in all networks.
We set two kind of parameters with different topological mixing parameter while leaving the mixing parameter $\mu_w$ as the independent variable. The network sizes are fixed to 150, the average degree is 15, and the maximum degree is 30, and all other parameters are identical to that in \cite{lancichinetti2009community}. The results are illustrated in Fig. \ref{figure_small_weighted}.
For algorithms that have an objective function, each point in Fig. \ref{figure_small_weighted} shows the average rrNMI over the results with highest objective function values in 10 runs on each of 30 networks.

When $\mu_w$ is 1.0, communities in planted partition have total intra-community edge weights close to zero. Both LPA and the coordinate ascent method for GSBM-P fail to obtain partitions similar to the planted partition. WSBM can group vertices with similar connectivity pattern, and do discover some communities with total intra-community edge weights close to zero. Thus, rrNMI is much larger than 0 for WSBM when $\mu_w$ is 1.0.

The variational Bayes algorithm for WSBM is actually not efficient. For example, it often reaches a poor local optimum. Only sometimes, it reaches a better result with higher marginal likelihood, which may not presents in 10 runs. Sometimes, the variational Bayes algorithm even is not converged in 200 iterations. From Fig. \ref{figure_small_weighted}, WSBM fails to give a good partition even when mixing parameter $\mu_w$ is small, which may due to the lack of degree correction for modeling edge weights. Moreover, the time complexity of each iteration of naive variational Bayes algorithm for WSBM is $\mathcal{O}(mK^2)$, where m denotes the total number of edges in networks and K denotes the number of communities, which limits its application in very large networks.
The coordinate ascent method for GSBM-P performs much better, and is comparable to LPA. In Fig. \ref{figure_7}, when mixing parameter $\mu_w$ is small, it recovers the planted partition.
For OSLOM, it is sensible to topological mixing parameter $\mu_t$. In figure \ref{figure_8}, the performance of OSLOM is unsatisfactory, and OSLOM identifies each vertex as a community when $\mu_w$ is larger than 0.35. In figure \ref{figure_7}, OSLOM returns same communities as planted communities in some tested benchmark networks with small $\mu_w$, but different communities in other tested benchmark networks with small $\mu_w$.

\begin{figure*}[!htbp]
\centering
\subfigure[]{
\begin{minipage}[t]{0.45\linewidth}
\centering
\includegraphics[width=3in]{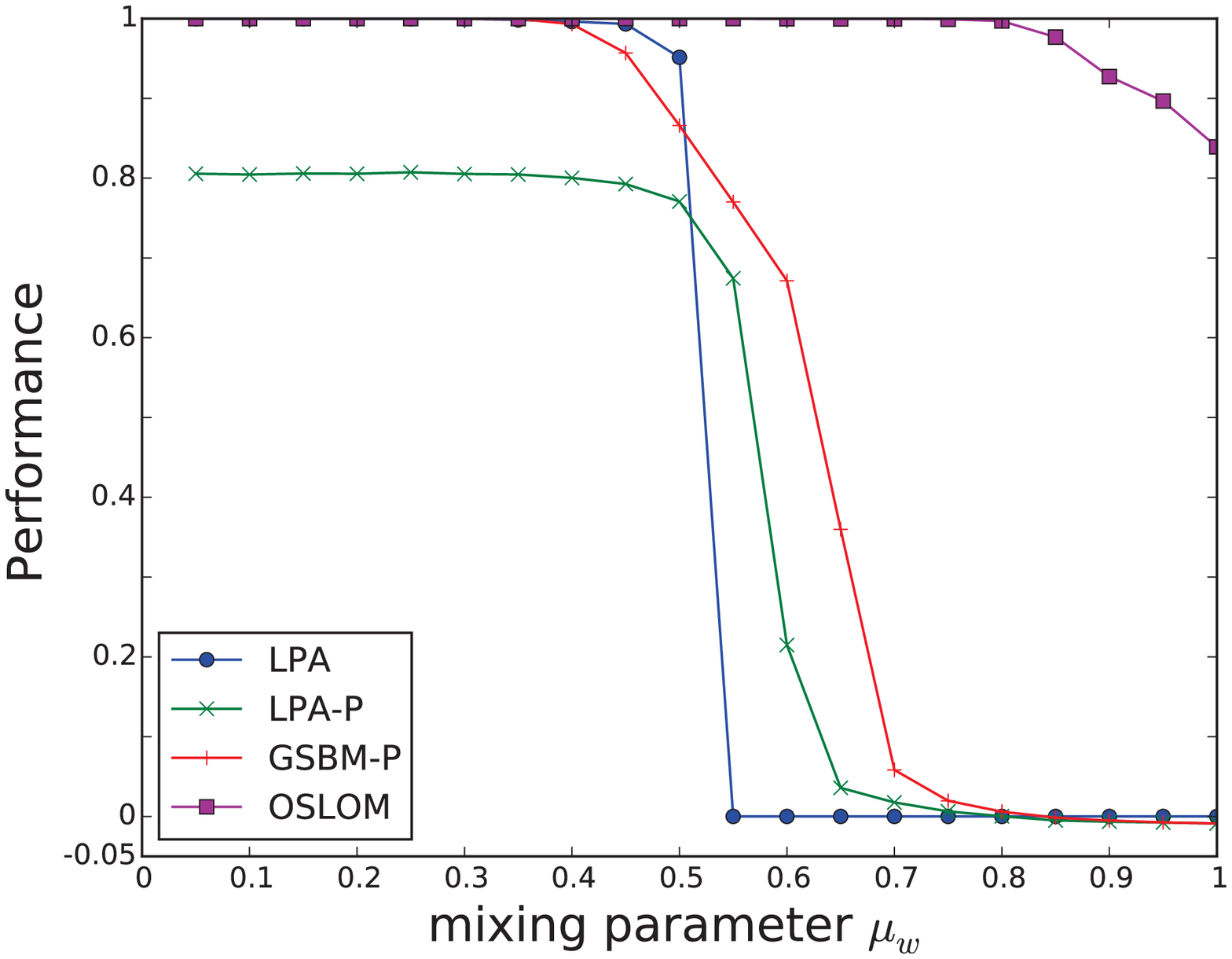}
\label{figure_3}
\end{minipage}
}
\subfigure[]{
\begin{minipage}[t]{0.45\linewidth}
\centering
\includegraphics[width=3in]{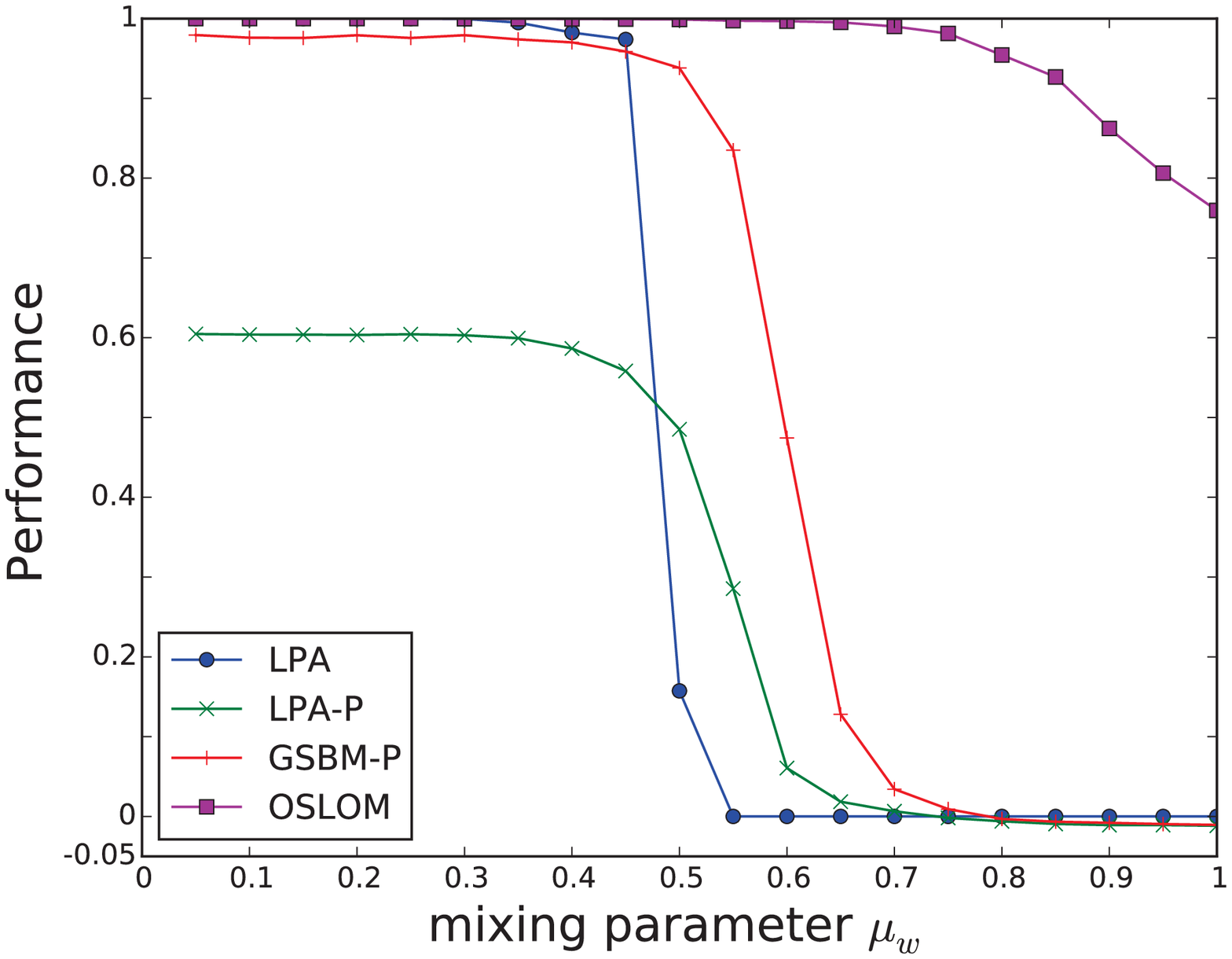}
\label{figure_4}
\end{minipage}
}
\subfigure[]{
\begin{minipage}[b]{0.45\linewidth}
\centering
\includegraphics[width=3in]{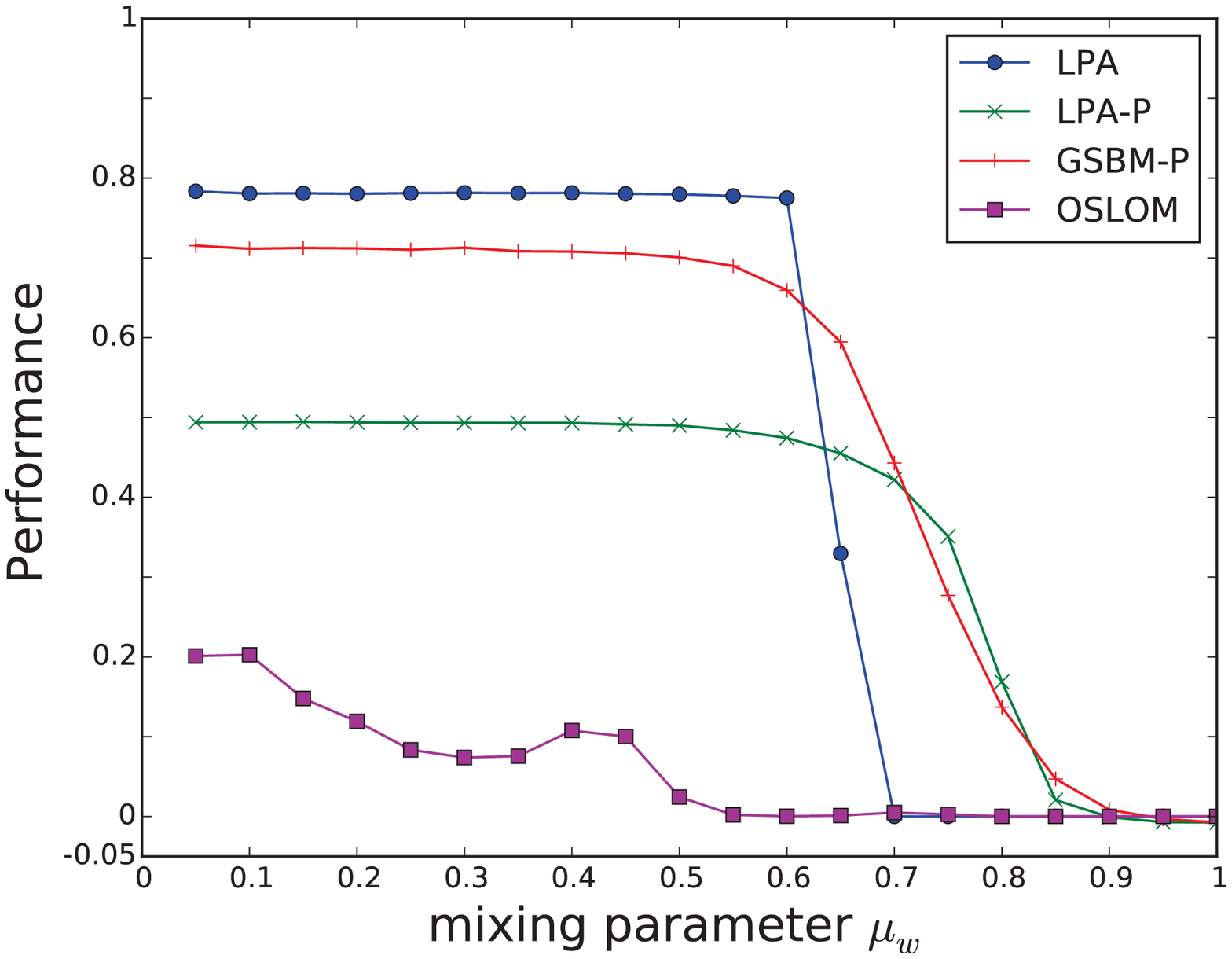}
\label{figure_5}
\end{minipage}
}
\subfigure[]{
\begin{minipage}[b]{0.45\linewidth}
\centering
\includegraphics[width=3in]{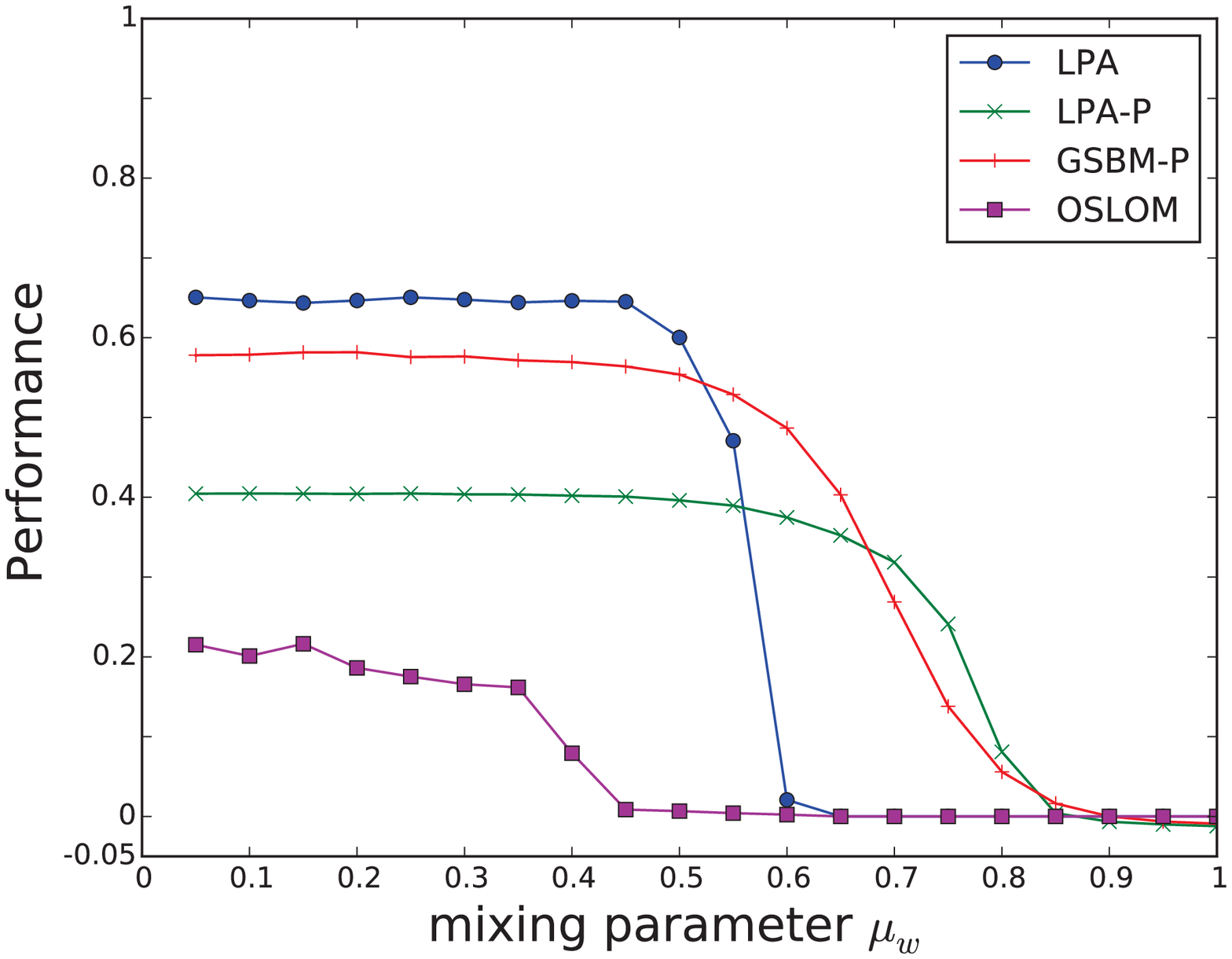}
\label{figure_6}
\end{minipage}
}
\caption{Performance with respect to the mixing parameter $\mu_w$ for different algorithms on four weighted test benchmarks \cite{lancichinetti2009benchmarks}. We use ratio of relative normalized mutual information (Eq. (\ref{rrNMI})) for LPA, LPA-P and GSBM-P. We use adjusted NMI \cite{lancichinetti2009detecting} for OSLOM. Each data point is an average over 30 networks generated by same parameters. In four subfigures, the size of networks is 1000, the average degree is 20, and the maximum degree is 50. In (a), the range of community size is $C=[10,50]$ and the topological mixing parameter $\mu_t$ is 0.5. In (b), the range of community size is $C=[20,100]$ and the topological mixing parameter $\mu_t$ is 0.5. In (c), the range of community size is $C=[10,50]$ and the topological mixing parameter $\mu_t$ is 0.8. In (d), the range of community size is $C=[20,100]$ and the topological mixing parameter $\mu_t$ is 0.8.}
\label{figure_weighted}
\end{figure*}

We show the performance of our method in weighted networks with 1000 vertices. We set four kind of parameters with different range of community size and different topological mixing parameter according to \cite{lancichinetti2009community} while leaving the mixing parameter $\mu_w$ as the independent variable. The results are illustrated in Fig. \ref{figure_weighted}. Here we omit the presentation of WSBM, because it takes too much time to run on these networks.

In all of the four groups of weighted benchmark networks, the coordinate ascent method for GSBM-P recovers the planted partition(see Fig. \ref{figure_3}) or achieves the results whose rrNMI are only slightly lower than that of LPA (see Fig. \ref{figure_4}, Fig. \ref{figure_5} and Fig. \ref{figure_6}). When mixing parameter $\mu_w$ is slightly larger, our method outperforms LPA.
In Fig. \ref{figure_5} and Fig. \ref{figure_6}, LPA discovers much more communities than the planted partition even when the mixing parameter $\mu_w$ is small. It implies that the communities are not well connected inside themselves.
The performance of our method is again comparable to LPA on those benchmark networks, while LPA-P is not satisfying.
OSLOM performs really well in benchmark networks when $\mu_t=0.5$, but again fails when $\mu_t=0.8$.

\section{Conclusions and Future Work}\label{SectionConclusions}

In this paper, we proposed the Gaussian stochastic blockmodel with node preference (GSBM-P). The maximum likelihood estimation of GSBM-P is proved to be equivalent to maximizing the objective function below:
\[ Q_\mathrm{GSBM-P} = \sum_z \lambda_z ^2 \;, \]
where $\lambda_z$ is the principal eigenvalue of the adjacency submatrix in community $z$.

We then proved that our coordinate ascent method optimizes our objective function in a similar way to label propagation with node preference.
We demonstrated that the vector composed by node preferences of vertices inside community $z$ is the product of $\sqrt{\lambda_z}$ and the intra-community eigenvector centrality,
i.e. the $L_2$-normalized dominant eigenvector of the adjacency matrix inside the community.

Experiments showed that the coordinate ascent method for Gaussian stochastic blockmodel with node preference worked well in most cases. It outperforms the variational Bayes method for weighted stochastic blockmodel and label propagation with intra-community RandomWalk as node preference in the aspect of community detection. In unweighted networks, the coordinate ascent method for GSBM-P is superior to LPA, and achieves similar performance to OSLOM. In weighted networks, the performance of our method is comparable to LPA.

\v{S}ubelj and Bajec \cite{vsubelj2011robust} have proposed that the iterative order implicitly contains the propagation strength. In the future, we may explore the suitable iterative order for our coordinate ascent method.
Moreover, coordinate ascent method is not the only way to optimize our objective function. We may explore different optimization algorithms in the future. We make two strong assumptions for our model, such that it is related to label propagation algorithms. However, these assumptions may be strict. The first one is that we fixed the expected edge weights between distinct communities to be zero. In the future, we may try to leave this small expected inter-community edge weight as a tunable parameter. The second one is that we treat the block indicators as parameters. Possible future work also includes placing a prior on block indicators and adopts more advanced approximation inference methods such as variational Expectation Maximization.

\section{Acknowledgement}

We thank Christopher Aicher for providing the implementation of weighted stochastic blockmodel. We thank Andrea Lancichinetti, Filippo Radicchi, Jos\'e Javier Ramasco and Santo Fortunato for providing the implementation of order statistics local optimization method. We thank Pan Zhang for providing the implementation of relative normalized mutual information.

\section{References}
\bibliography{GSBM}

\providecommand{\newblock}{}
\begin{thebibliography}{10}
\expandafter\ifx\csname url\endcsname\relax
  \def\url#1{{\tt #1}}\fi
\expandafter\ifx\csname urlprefix\endcsname\relax\def\urlprefix{URL }\fi
\providecommand{\eprint}[2][]{\url{#2}}

\bibitem{girvan2002community}
Girvan M and Newman M~E 2002 {\em Proceedings of the National Academy of
  Sciences\/} {\bf 99} 7821--7826

\bibitem{newman2004finding}
Newman M~E~J and Girvan M 2004 {\em Physical Review E\/} {\bf 69} 026113

\bibitem{raghavan2007near}
Raghavan U~N, Albert R and Kumara S 2007 {\em Physical Review E\/} {\bf 76}
  036106

\bibitem{vsubelj2011robust}
{\v{S}}ubelj L and Bajec M 2011 {\em The European Physical Journal B-Condensed
  Matter and Complex Systems\/} {\bf 81} 353--362

\bibitem{barber2009detecting}
Barber M~J and Clark J~W 2009 {\em Physical Review E\/} {\bf 80} 026129

\bibitem{leung2009towards}
Leung I~X, Hui P, Lio P and Crowcroft J 2009 {\em Physical Review E\/} {\bf 79}
  066107

\bibitem{vsubelj2011unfolding}
{\v{S}}ubelj L and Bajec M 2011 {\em Physical Review E\/} {\bf 83} 036103

\bibitem{liu2010advanced}
Liu X and Murata T 2010 {\em Physica A: Statistical Mechanics and its
  Applications\/} {\bf 389} 1493--1500

\bibitem{schuetz2008efficient}
Schuetz P and Caflisch A 2008 {\em Physical Review E\/} {\bf 77} 046112

\bibitem{brin1998anatomy}
Brin S and Page L 1998 {\em Computer networks and ISDN systems\/} {\bf 30}
  107--117

\bibitem{holland1983stochastic}
Holland P~W, Laskey K~B and Leinhardt S 1983 {\em Social networks\/} {\bf 5}
  109--137

\bibitem{hofman2008bayesian}
Hofman J~M and Wiggins C~H 2008 {\em Physical Review Letters\/} {\bf 100}
  258701

\bibitem{karrer2011stochastic}
Karrer B and Newman M~E 2011 {\em Physical Review E\/} {\bf 83} 016107

\bibitem{airoldi2008mixed}
Airoldi E~M, Blei D~M, Fienberg S~E and Xing E~P 2008 {\em Journal of Machine
  Learning Research\/} {\bf 9} 3

\bibitem{aicher2014learning}
Aicher C, Jacobs A~Z and Clauset A 2014 {\em arXiv preprint arXiv:1404.0431\/}

\bibitem{gopalan2013modeling}
Gopalan P, Wang C and Blei D 2013 Modeling overlapping communities with node
  popularities {\em Advances in Neural Information Processing Systems\/} pp
  2850--2858

\bibitem{aicher2013adapting}
Aicher C, Jacobs A~Z and Clauset A 2013 {\em arXiv preprint arXiv:1305.5782\/}

\bibitem{tibely2008equivalence}
Tib{\'e}ly G and Kert{\'e}sz J 2008 {\em Physica A: Statistical Mechanics and
  its Applications\/} {\bf 387} 4982--4984

\bibitem{bonacich1987power}
Bonacich P 1987 {\em American journal of sociology\/}  1170--1182

\bibitem{lancichinetti2011finding}
Lancichinetti A, Radicchi F, Ramasco J~J and Fortunato S 2011 {\em PloS one\/}
  {\bf 6} e18961

\bibitem{kleinberg1999authoritative}
Kleinberg J~M 1999 {\em Journal of the ACM\/} {\bf 46} 604--632

\bibitem{he2014construction}
He S, Zou X, Xiao L and Hu J 2014 Construction of diachronic ontologies from
  people¡¯s daily of fifty years {\em Language Resources and Evaluation
  Conference\/} pp 3258--3263

\bibitem{traag2011narrow}
Traag V~A, Van~Dooren P and Nesterov Y 2011 {\em Physical Review E\/} {\bf 84}
  016114

\bibitem{zachary1977information}
Zachary W 1977 {\em Journal of anthropological research\/} {\bf 33} 452--473

\bibitem{adamic2005political}
Adamic L~A and Glance N 2005 The political blogosphere and the 2004 us
  election: divided they blog {\em Proceedings of the 3rd international
  workshop on Link discovery\/} (ACM) pp 36--43

\bibitem{lancichinetti2008benchmark}
Lancichinetti A, Fortunato S and Radicchi F 2008 {\em Physical Review E\/} {\bf
  78} 046110

\bibitem{lancichinetti2009benchmarks}
Lancichinetti A and Fortunato S 2009 {\em Physical Review E\/} {\bf 80} 016118

\bibitem{erdos1959random}
Erd\"os P and R\'enyi A 1959 {\em Publ. Math. Debrecen\/} {\bf 6} 290--297

\bibitem{witten2005data}
Witten I~H and Frank E 2005 {\em Data Mining: Practical machine learning tools
  and techniques\/} (Morgan Kaufmann)

\bibitem{zhang2015revisit}
Zhang P 2015 {\em arXiv preprint arXiv:1501.03844\/}

\bibitem{lancichinetti2009detecting}
Lancichinetti A, Fortunato S and Kert{\'e}sz J 2009 {\em New Journal of
  Physics\/} {\bf 11} 033015

\bibitem{agarwal2008modularity}
Agarwal G and Kempe D 2008 {\em The European Physical Journal B-Condensed
  Matter and Complex Systems\/} {\bf 66} 409--418

\bibitem{koutsourelakis2008finding}
Koutsourelakis P~S and Eliassi-Rad T 2008 Finding mixed-memberships in social
  networks. {\em AAAI Spring Symposium: Social Information Processing\/} pp
  48--53

\bibitem{lancichinetti2009community}
Lancichinetti A and Fortunato S 2009 {\em Physical Review E\/} {\bf 80} 056117

\bibitem{fortunato2007resolution}
Fortunato S and Barthelemy M 2007 {\em Proceedings of the National Academy of
  Sciences\/} {\bf 104} 36--41

\bibitem{vsubelj2011generalized}
{\v{S}}ubelj L and Bajec M 2011 {\em arXiv preprint arXiv:1110.2711\/}

\bibitem{radicchi2004defining}
Radicchi F, Castellano C, Cecconi F, Loreto V and Parisi D 2004 {\em
  Proceedings of the National Academy of Sciences\/} {\bf 101} 2658--2663

\end{thebibliography}

\end{document}